\documentclass[a4paper,fleqn,usenatbib]{mnras}

\usepackage{graphicx}	
\usepackage{amsmath}	
\usepackage{amssymb}	
\usepackage{subeqnarray}
\usepackage{hyperref}
\usepackage[para]{threeparttable} 
\usepackage[usenames, dvipsnames]{color}

\def\vsini{V_\mathrm{eq} \sin i} 
\def\kms{\mathrm{km.s}^{-1}}
\def\phidiff{\phi_\mathrm{diff}}
\def\Rsun{\mathrm{R}_{\sun}}

\def\Msun{\mathrm{M}_{\sun}}
\def\Tmean{\overline{T}_\mathrm{eff}}

\title[Limits of the photocentre on fast rotators]{Application limit of the photocentre displacement to fundamental stellar parameters of fast rotators\\ - Illustration on the edge-on fast rotator Regulus}

\author[M. Hadjara et al.]{
M. Hadjara,$^{1,2,3,4}$\thanks{E-mail: Massinissa.Hadjara@oca.eu (MH), Romain.Petrov@unice.fr (RP) and sjankov@aob.rs (SJ)}
R. G. Petrov,$^{5}$
S. Jankov,$^{6}$
P. Cruzal\`ebes,$^{5}$
A. Boskri,$^{5,7}$
A. Spang,$^{5}$
\newauthor
S. Lagarde,$^{5}$
J. He,$^{8,1,2}$
X. Chen,$^9$
C. Nitschelm,$^{10}$
E. S. G. de Almeida,$^{5}$
G. Pereira,$^{11}$
\newauthor
E. A. Michael,$^{11}$
Q. Gao,$^{1}$
W. Wang,$^{1,12}$
I. Reyes,$^{11}$
C. Arcos,$^{13}$
I. Araya,$^{14}$
M. Cur\'e$^{13}$
\\
$^{1}$Chinese Academy of Sciences South America Center for Astronomy, National Astronomical Observatories, CAS, Beijing 100101, China\\
$^{2}$Departamento de Astronomía, Universidad de Chile, Casilla 36-D, Santiago, Chile\\
$^{3}$Instituto de Astronom{\'i}a, Universidad Cat{\'o}lica del Norte, Av. Angamos 0610 Antofagasta, Chile\\
$^{4}$Centre de Recherche en Astronomie, Astrophysique et G\'{e}ophysique (CRAAG), Route de l'Observatoire, B.P. 63, Bouzareah, 16340,\\ Alger, Algeria\\
$^{5}$Universit\'e C\^ote d’Azur (UCA), Centre National de la Recherche Scientifique (CNRS), Observatoire de la C\^ote d’Azur (OCA),\\
 Laboratoire J. L. Lagrange, UMR 7293,Campus Valrose, 06108 Nice Cedex 2, France\\
$^{6}$Astronomical Observatory, Volgina 7, P.O.Box 74 11060 Belgrade, Serbia\\
$^{7}$LPHEA Laboratory, Oukaimeden Observatory, Cadi Ayyad University/FSSM, BP 2390 Marrakesh, Morocco,\\
$^{8}$Yunnan Observatories, Chinese Academy of Sciences, 396 Yangfangwang, Guandu District, Kunming, 650216, China,\\
$^{9}$Optical Interferometry Group, Shanghai Astronomical Observatory(SHAO), Chinese Academy Sciences(CAS) , Shanghai, 200030, China,\\
$^{10}$Centro de Astronom{\'i}a (CITEVA), Universidad de Antofagasta, Avenida Angamos 601, Antofagasta 1270300, Chile\\
$^{11}$Radio Astronomical Instrumentation Group (RAIG), Terahertz- and Astro-Photonics Laboratory, Departamento de Ingenier\'ia El\'ectrica, \\ Universidad de Chile, Avenida Tupper 2007, Santiago, Chile\\
$^{12}$CAS Key Laboratory of Optical Astronomy, National Astronomical Observatories, Chinese Academy of Sciences, Beijing 100101, China\\
$^{13}$Instituto de F\'isica y Astronom\'ia, Facultad de Ciencias, Universidad de Valpara\'iso, Av. Gran Breta\~na 1111, Valpara\'iso, Chile\\
$^{14}$Centro de Investigación DAiTA Lab, Facultad de Estudios Interdisciplinarios, Universidad Mayor, Alonso de Córdova 5495, Santiago, Chile}
\date{Accepted 2022 January 11. Received 2022 January 11; in original form 2021 August 16}

\pubyear{2022}

\begin{document}
\label{firstpage}
\pagerange{\pageref{firstpage}--\pageref{lastpage}}
\maketitle

\begin{abstract}
Differential Interferometry allows to obtain the differential visibility and phase, in addition to the spectrum. The differential phase contains important information about the structure and motion of stellar photosphere such as stellar spots and non-radial pulsations, and particularly the rotation.
Thus, this interferometric observable strongly helps to constrain the stellar fundamental parameters of fast rotators. The spectro-astrometry mainly uses the photocentre displacements, which is a first approximation of the differential phase, and is applicable only for unresolved or marginally objects.\\
We study here the sensitivity of relevant stellar parameters to the simulated photocentres using the SCIROCCO code: a semi-analytical algorithm dedicated to fast rotators, applied to two theoretical modeling stars based on Achernar and Regulus, in order to classify the importance of these parameters and their impact on the modeling. We compare our simulations with published VLTI/AMBER data.\\
This current work sets the limits of application  of photocentre displacements to fast rotators, and under which conditions we can use the photocentres and/or the differential phase, through a pre-established physical criterion. To validate our theoretical study, we apply our method of analysis on observed data of the edge-on fast rotator Regulus.
For unresolved targets, with a visibility $V\sim 1$, the photocentre can constrain the main stellar fundamental parameters of fast rotators, whereas from marginally resolved objects ($0.8 \leq V < 1$), mainly the rotation axis position angle ($\rm PA_{\rm rot}$) can be directly deduced from the vectorial photocentre displacement, which is very important for young cluster studies.
\end{abstract}


\section{Introduction }
\label{introduction}

\subsection{Fundamental parameters of stars}
\label{Fund_param_stars}
Measuring the fundamental parameters of stars, such as diameter, mass, rotation, effective temperature or age remains crucial for stellar physics. In particular this is necessary to properly characterize the host stars and discuss the evolution of the thousands of extrasolar planets that have been discovered from radial-velocimetry and transit observations. It is also important to combine the fundamental parameters with the stellar activity and asteroseismology through observational breakthroughs offered by the high-performance interferometers, spectrographs and photometers. 
The angular resolution provided by optical long baseline interferometry is crucial to constrain most of these fundamental parameters, including the proper separation between local and global velocity fields \citep[][and references therein]{2011SerAJ.183....1J}.

\subsection{Rapid rotators}
\label{Fast_rotators}
In that context, rapid rotators are of specific interest. Rotation velocities close to the critical velocity are keys to understand mass loss and stellar winds. The geometrical flattening, coupled with gravity darkening \citep{1924MNRAS..84..665V} -known as the von Zeipel effect- and the resulting lower luminosity and radiation pressure at the stellar equator, is also a key to shape the polar components of the stellar wind \citep{2006A&A...453.1059K}.
The mechanisms amplified by rapid rotation, such as meridional circulation or turbulence, may affect the internal structure of the star and its evolution \citep{2009LNP...765..139M}. The gravity darkening has a profound impact on the physics of the stars, with important observational consequences. For example, the models from \citet{1977ApJS...34...41C} indicate a two-component spectral energy distribution (SED) for these stars, with an  infrared excess due to gravity darkening. Therefore, it is not easy to include these stars in a single spectral class, since the observed SED depends on their rotational velocity and inclination angle.

The most important parameters of a rapid rotator are : 
\begin{itemize}
\item The equatorial and polar radii $R_{\rm eq}$ and $R_{\rm pol}$, respectively, which contains information about the size of the star;
\item The stellar flattening $R_{\rm eq}/R_{\rm pol}$, which is a key constraint on the gravity darkening effect and the differential rotation;
\item The equatorial velocity $V_{\rm eq}$ , which is connected to the kinematics of the stellar photosphere; 
\item The inclination angle $i$, which is the angle between the observation line of sight and the stellar rotation axis. This parameter provides the true spectral type due to the gravity darkening effect and influencing the nature of Be stars \citep[see.g.][]{2003PASP..115.1153P};
\item The gravitational darkening is mostly described by the coefficient $\beta$ \citep[e.g.][]{2014A&A...569A..45H}. This stellar parameter depicts the difference of temperature, luminosity and surface gravity between the equator and the poles;
\item The rotation axis position angle $\rm PA_{\rm rot}$, which is the angle of the projected rotation axis on the sky, measured from north to east. It can give, together with the proper motion of the star, information on the protostellar cloud dynamics. The knowledge of this parameter is very important in multiple systems, or in the presence of circumstellar material to constrain the evolution of the system.
\end{itemize}

\subsection{Rapidly rotating star modeling for Differential Interferometry}
\label{HAS-DI}

Differential Interferometry (DI) is a technique based on information taken at different wavelengths. In the case of long-baseline interferometry, it uses spectrally dispersed fringes. This technique offers two major advantages: i) getting information beyond the instrumental angular resolution limit, and ii) measuring simultaneously spatial, spectroscopic and kinematics properties of the stellar surface, with angular diameter down to $0.1\,\rm mas$, which is for example 100 parsec for Sun-like stars.
Inspired by the early works of \cite{1975ApJ...196L..71L} and \cite{1982AcOpt..29..361B} proposed the Differential Speckle Interferometry technique, which uses the chromatic displacement of the speckle photocentre, given by the first-order term of the phase according to the MacLaurin series \citep{2001A&A...377..721J}. This technique has been extended to a wider range of wavelengths and applied to long-baseline interferometry by \cite{1988ESOC...29..235P, 1989dli..conf..249P} who established the fundamentals of the DI technique. For the first time, this allowed to separate spatial and spectral information of two stellar components forming the binary Capella \citep[][]{1992ASPC...32..477P} and to measure the stellar rotation of the slow rotator Aldebaran using the photocentre displacement \citep{sl94}.
The first theoretical general study of the DI technique was done by \cite{1995A&AS..109..401C} to estimate the angular diameters, the rotation velocities and the position angles of the rotation axis of single stars, as well as the angular separations and the radial velocity differences of close binary systems.
The DI offers the possibility to constrain these parameters simultaneously \citep{1995A&AS..109..401C}.

It is imperative to mention the spectro-astrometry in this paper, because it is a method of photocentre measurement \citep{2018Natur.563..657G,2018MNRAS.480.1263H,2009A&A...498L..41L}. The potential of this technique with very large telescopes is obvious, because the precision on the measurement of a photocentre improves as $B\times D$ (the interferometric baseline length times the telescope diameter) in the visible and as $B\times D^2$ in the thermal IR or when we are limited by the detector noise \citep{1989dli..conf..249P}.
Several E-ELT instruments will provide spectro-astrometry mode like METIS \citep{2008SPIE.7014E..1NB} or possibly HIRES \citep{2013arXiv1310.3163M}. As well as two possible future VLTI visitor instruments, in J-band are in being studied/carried out currently, namely: BIFROST \citep[][with a very rich spectro-astrometry program]{2018tcl..confE..29K} and VERMILION (scientific white paper in prep.), with its new generation fringe tracker \citep{2019vltt.confE..37P}. The spectral resolution is also very important, for further study of the spectroscopic lines. Currently, at the VLTI, GRAVITY \citep{2017A&A...602A..94G} offers a spectral resolution $R\sim 4500$ in K-band, where AMBER \citep{2007A&A...464....1P}, before its decommissioning, was offering $R\sim 12000$ at the same band. BIFROST and VERMILION expect to offer $R\sim 20000$ to $30000$.

Measuring the stellar rotation is essential : ignoring it can induce spurious conclusions, such as a mis-classification of the spectral type when the star rotates rapidly around its rotation axis, which has a strong influence on the emitted flux \citep[e.g.][]{1972A&A....21..279M} and on the gravity darkening \citep{1924MNRAS..84..684V}, for example the reference star Vega, has been discovered to be a pole-on rapid rotator \citep{2006Natur.440..896P, 2006ApJ...645..664A}.

Numerical models of fast rotators, possibly including their close stellar environment (CSE), are rather rare, very few of them being dedicated to optical interferometry : 
\begin{itemize}
\item For the fast rotators only, let us mention: CHARRON \citep{2012sf2a.conf..321D} and ESTER \citep{2016JCoPh.318..277R,2013A&A...552A..35E},
\item For the CSE only, lets us mention : SIMECA \citep{2008EAS....28..135S}, BE-DISK \citep{2009ApJ...699.1973S}, and the numerical model of \cite{2012ApJ...744...19K} which studies the photocentre displacement of a Be envelope in local thermodynamic equilibrium from infrared interferometric measurements,
\item For both the rotating stars and their disks, let us mention HDUST \citep{2006ApJ...639.1081C}, a radiative transfer code which produces spectra and intensity maps in natural and/or polarized light for CSEs of massive stars including gas and dust.
\end{itemize}

In this paper, we further describe in detail the methodology introduced by \cite{2018MNRAS.480.1263H}. We study and discuss the limit and dependence of the photocentre defined by its angular coordinates $(\epsilon_\alpha,\epsilon_\delta)$ that we deduce from the differential phase $\phidiff$, to the relevant physical parameters of our numerical model SCIROCCO \citep{2012sf2a.conf..533H, 2013EAS....59..131H, 2014A&A...569A..45H, 2018MNRAS.480.1263H}, using high spectral resolution interferometric data such as those obtained with the VLTI-AMBER in the K band (R$\sim$12000) for a given star with any inclination angle ($0^\circ \leq i \leq 90^\circ$). As an example, we chose to use a reference modeling star similar to Achernar, a Be star of spectral type B6V and an inclination angle $i$ of about $60^\circ$, but with a rotation axis position angle $\rm PA_{\rm rot}=0^\circ$ to simplify \citep[as it is shown in][]{2012A&A...545A.130D}.
We show that, in the case of fast rotators, the polar photocentre displacement $(\epsilon_{\rm pol})$ is sensitive to the gravity darkening, where the equatorial one $(\epsilon_{\rm eq})$ is sensitive to the stellar rotational velocity.
This study allows us to determine the limits of application of photocentre displacements to fast rotators, and under which conditions we can use the photocentres and/or the differential phase, through a pre-established physical criterion. Then, we study and discuss the sensitivity of $(\epsilon_\alpha,\epsilon_\delta)$ to relevant stellar parameters for an edge-on star (as an example, a reference modeling star similar to Regulus, which is a B sub-giant star of spectral type B8IV), with an $\rm PA_{\rm rot}\neq0^\circ$ and $i\approx90^\circ$ \citep[][]{2018MNRAS.480.1263H}, which allows us to study the impact of the gravity darkening coefficient $\beta$ on $(\epsilon_\alpha,\epsilon_\delta)$. The present paper is organized as follows:
\begin{itemize}
\item In Sect.~\ref{phidiff}, we introduce the differential phase ($\phidiff$).
\item In Sect.~\ref{photocentres}, we show how and within which limits the vectorial right-ascension/declination photocentre $(\epsilon_\alpha,\epsilon_\delta)$ can be deduced from the differential phase $\phidiff$.
\item In Sect.~\ref{phot-sensitivity}, we study the influence of relevant physical parameters on the photocentre displacement and on the flux, using our model of an Achernar-like star with an arbitrary inclination.
\item In Sect.~\ref{AnnexC}, we show the application of the photocentre displacement method to real data of Regulus.
\item In Sect.~\ref{sensitiv_eps}, we discuss the photocentre displacement sensitivity to the key stellar parameters; $R_{\rm eq}$, $V_{\rm eq}$, $i$, $\rm PA_{\rm rot}$, and $\beta$ on our second modeling reference star, through the best results obtained from Sect.~\ref{AnnexC}.
\item In Sect.~\ref{conclusions}, we analyze and discuss our results.
\item In Appendix.~\ref{AnnexAlpha}, we explain a few important details about the modeling for fast rotators with SCIROCCO.
\end{itemize}

\section{Simulated differential phase for rapidly rotating stars}
\label{phidiff}

The differential phase $\phidiff$ measured for the on-sky-projected interferometric baseline $\vec{B}(u,v)$, as function of spatial frequencies $u$ and $v$, at the reference wavelength $\lambda_0$ taken in the continuum, is defined as :
\begin{equation}
\phidiff(u,v,\lambda)=arg\left(V(u,v,\lambda)\right)-arg\left(V(u,v,\lambda_0)\right), 
\label{eq:1}
\end{equation}
where $V$ is the object complex visibility. Although the phase $arg\left(V(u,v,\lambda)\right)$ is not recovering well by interferometer due to atmospheric turbulence and lack of absolute reference, the differential phase $\phidiff$ is well retrieved.
This quantity is directly connected to the chromatic displacement of the photocentre vector according to :
\begin{equation}
\epsilon(\lambda)=-\frac{\phidiff(u, v, \lambda)}{2\pi\sqrt{u^2+v^2}},  
\label{eq:2}
\end{equation}
given by the first-order approximation of the McLaurin series of the phase \citep[the demonstration, as well as the validity limits, of this development have been handled by][]{2001A&A...377..721J}.

\begin{figure}
\centering
 \includegraphics[width=1.\hsize,draft=false]{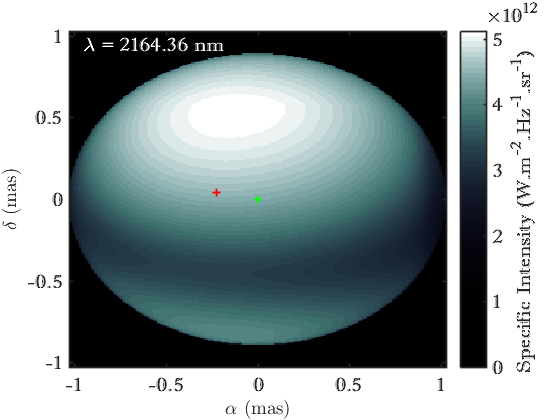}
 \includegraphics[width=1.\hsize,draft=false]{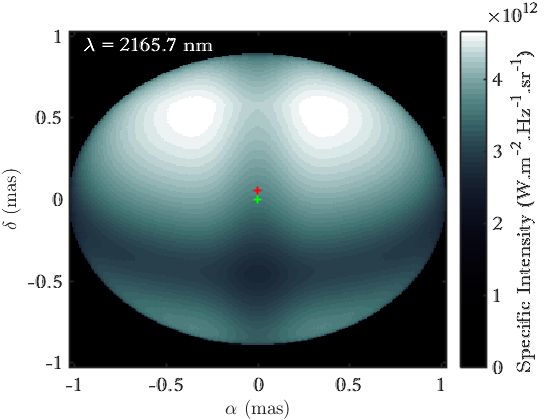}
 \includegraphics[width=1.\hsize,draft=false]{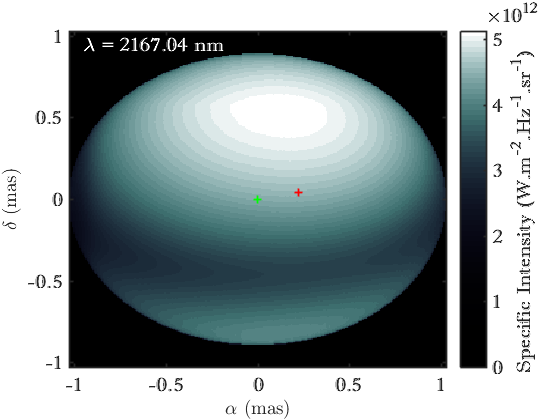}
 \caption{Monochromatic intensity maps simulated by SCIROCCO, with the physical parameters of the reference model similar to Achernar (see text for details), for 3 different wavelengths. The red crosses show the global/astrometric photocentre position (explained in Sect.~\ref{photocentres}) corresponding to each wavelength, while the green crosses depict the photocentre position in the continuum.} \label{sim_Ach_Scirocco}
\end{figure}

In order to study the sensitivity $\epsilon_\alpha$ and $\epsilon_\delta$ (that we deduce from $\phidiff$, and which are linked to the right ascension and the declination respectively) we use the numerical code SCIROCCO described in \cite{2014A&A...569A..45H}, to various stellar parameters, as we did previously \citep{2012A&A...545A.130D} for a reference modeling star, at any inclination angle ($0^\circ < i < 90^\circ$), similar to Achernar with the CHARRON code described in detail in \cite{2012sf2a.conf..321D, 2002A&A...393..345D}. Namely, here we use: $R_{eq} = 11\,\Rsun$, $d = 50\,pc$, $\vsini = 250\,\kms$, $i = 60^\circ$, $M = 6.1\,\Msun$, $T_{\rm eff} = 15000\,\rm K$, $\rm PA_{\rm rot} = 0^\circ$ (northern direction), $\beta = 0.25$ (theoretical value for pure radiative stellar envelope according to \cite{1924MNRAS..84..684V}), without taking into account the differential rotation. In addition, we take into account the limb darkening effects to produce line profiles with KURUCZ/SYNSPEC stellar atmospheres model. These parameters correspond to $V_{\rm eq}$ equal to 90\% of the critical velocity $V_{\rm crit}$, $R_{eq}/R_p = 1.4$ and an equatorial angular diameter $\Theta = 2R_{eq}/d = 2\,\rm mas$. Figure \ref{sim_Ach_Scirocco} shows the monochromatics intensity maps for a given Doppler shift across 3 wavelengths around the absorption Br$\gamma$ line, adopting the stellar parameters given above \citep[more details about features of this kind of figures have been well described by][]{2014A&A...569A..45H,Massi2015,2018MNRAS.480.1263H}.

In the appendix \ref{AnnexAlpha}, we give more details about how the SCIROCCO code works, in particular on the flattened shape used of the fast rotating stars as well as on the line profile.

\section{FROM $\phidiff$ TO THE PHOTOCENTRE COORDINATES}
\label{photocentres}

\subsection{Deducing the photocentre coordinates}
\label{photocentres1}

\begin{figure}
\centering
 \includegraphics[width=1.\hsize,draft=false]{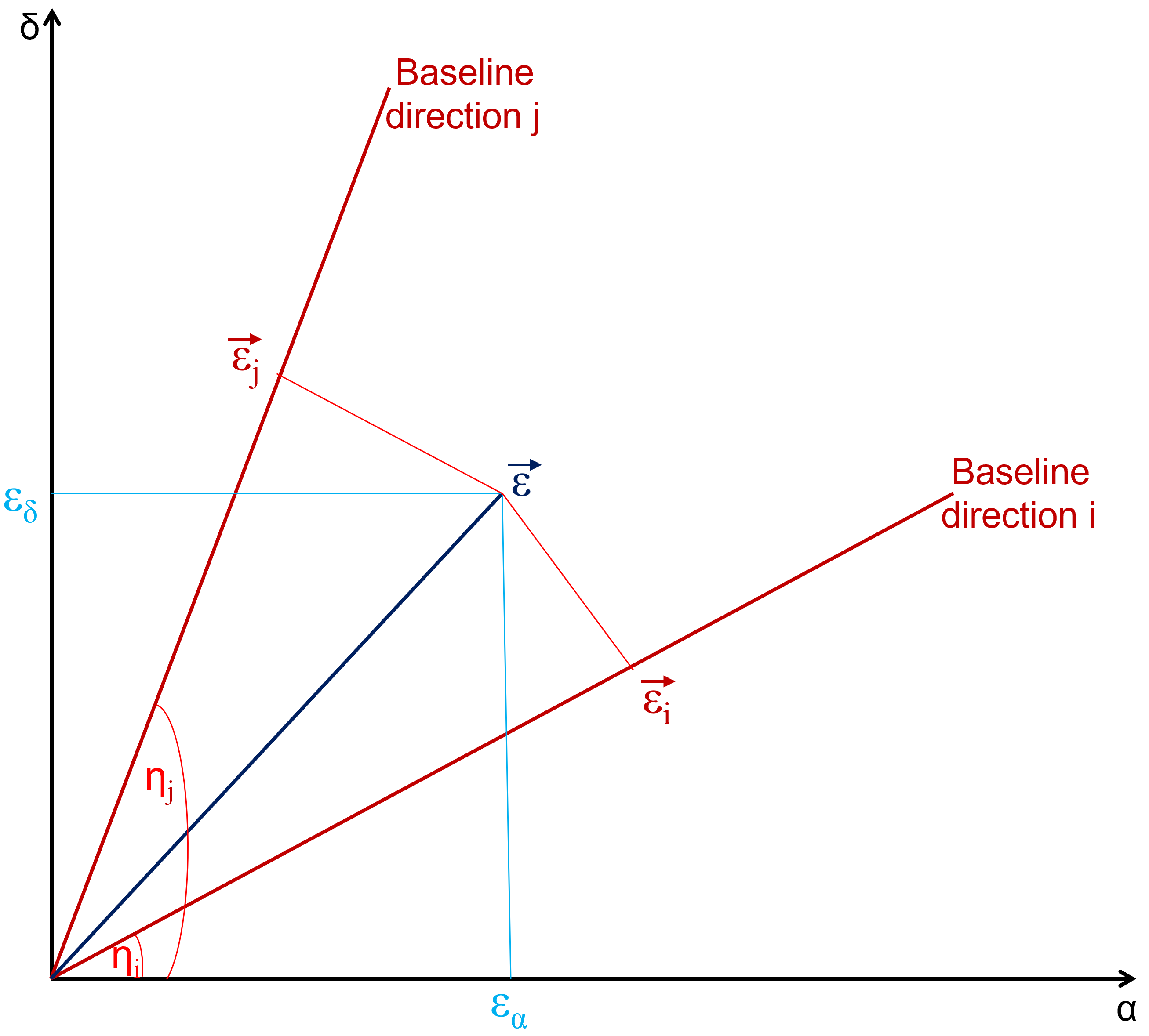}
 \caption{The photocentre vector $\vec{\epsilon}$ projected on two interferometric baseline directions $\eta_{\rm i}$ and $\eta_{\rm j}$ ($\epsilon_{\rm i}$ and $\epsilon_{\rm j}$) and projected on the right ascension and declination directions ($\epsilon_\alpha$ and $\epsilon_\delta$ respectively).}\label{phot-proj}
\end{figure}

We deduce the photocentre coordinates $(\epsilon_\alpha,\epsilon_\delta)$ from $\phidiff$ using elementary geometric projections as shown in Fig.~\ref{phot-proj}.

Let us call $\vec{\epsilon}_{\rm i}$ and $\vec{\epsilon}_{\rm j}$ the projections of the protocentre vector $\vec{\epsilon}$ on the directions of the baselines $\rm i$ and $\rm j$ respectively (with the projection angles $\eta_{\rm i}$ and $\eta_{\rm j}$ respectively).
We define the 2 ``pseudo-coordinates'' $\epsilon_{\alpha,ij}$ and $\epsilon_{\delta,ij}$ as :
\begin{subeqnarray}\label{epsxy_eq}
& \epsilon_{\alpha,\rm ij}=(\epsilon_{\rm i}\sin\eta_{\rm j}-\epsilon_{\rm j}\sin\eta_{\rm i})/\sin(\eta_{\rm j}-\eta_{\rm i}),  \\
& \epsilon_{\delta,\rm ij}=(\epsilon_{\rm j}\cos\eta_{\rm i}-\epsilon_{\rm i}\cos\eta_{\rm j})/\sin(\eta_{\rm j}-\eta_{\rm i}),
\end{subeqnarray}
where $\epsilon_{\rm i}$ and $\epsilon_{\rm j}$ are the moduli of the projected vectors $\vec{\epsilon}_{\rm i}$ and $\vec{\epsilon}_{\rm j}$, respectively. We can easily check these equations in Fig.~\ref{phot-proj}, where $(\eta_{\rm i}, \eta_{\rm j}) = (0, \pi/2)$ $\rightarrow$ $(\epsilon_\alpha, \epsilon_\delta)=(\epsilon_{\rm i}, \epsilon_{\rm j})$ and $(\eta_{\rm i}, \eta_{\rm j}) = (\pi/2, \pi)$ $\rightarrow$ $(\epsilon_\alpha, \epsilon_\delta)=(-\epsilon_{\rm j},\epsilon_{\rm i})$. Note that the angle $\eta$ is called the projection angle ($\rm PA$) of the baseline ($\rm PA=\eta$), and which is defined from north to east. Thus, from now onwards we will use the notation of $\rm PA$ instead of $\eta$, and $\Delta\eta=\left|\eta_{\rm i}-\eta_{\rm j}\right|$, will be simply called $\Delta \rm PA$.

Knowing that for any baseline $\rm n$ at each wavelength $\lambda$, the relation between the photocentre $\epsilon_{\rm n}(\lambda)$ and differential phase $\phi_{\rm n} (\lambda)$, which is only satisfied for marginally resolved and unresolved objects \citep[as demonstrated by][using the derivative and definite integral theorem to the Maclaurin expansion of the complex visibility function]{2001A&A...377..721J}, is given by:
\begin{equation}\label{phot-phi_eq}
\phi_{\rm n} = - \frac{2\pi}{\lambda} \vec{\epsilon}_{\rm n}(\lambda)\cdot\vec{B}_{\rm n}.
\end{equation}
If we use more than 2 independent baselines, we deduce the mean coordinates of the photocentre displacement $(\epsilon_\alpha,\epsilon_\delta)$ thanks to the weighted average, favoring the least noisy baselines, given by:
\begin{subeqnarray}\label{phot-tot_eq}
& \epsilon_\alpha (\lambda)=\sum_{\rm i,j>i}\frac{\epsilon_{\alpha,\rm ij}(\lambda)}{Var(\epsilon_{\alpha,\rm ij}(\lambda))}\big/\sum_{\rm i,j>i}\frac{1}{Var(\epsilon_{\alpha,\rm ij}(\lambda))},  \\
& \epsilon_\delta (\lambda)=\sum_{\rm i,j>i}\frac{\epsilon_{\delta,\rm ij}(\lambda)}{Var(\epsilon_{\delta,\rm ij}(\lambda))}\big/\sum_{\rm i,j>i}\frac{1}{Var(\epsilon_{\delta,\rm ij}(\lambda))},
\end{subeqnarray}
where $\textit{Var}$ is the measured variance.

The absolute photocentre coordinates $E_\alpha$ and $E_\delta$, measured separately by single-aperture astrometry, is directly deducted from the monochromatic intensity maps $I(x,y,\lambda)$, using the following formula:
\begin{subeqnarray}\label{phot-glob_eq}
& E_\alpha (\lambda)=\frac{\sum_\alpha\sum_\delta \alpha I(\alpha,\delta,\lambda)}{\sum_\alpha\sum_\delta I(\alpha,\delta,\lambda)},  \\
& E_{\delta}(\lambda)=\frac{\sum_\alpha\sum_\delta \delta I(\alpha,\delta,\lambda)}{\sum_\alpha\sum_\delta I(\alpha,\delta,\lambda)}.
\end{subeqnarray}

To keep the same convention as for $\phidiff$ in Eq.~\ref{eq:1}, which is zero on the continuum, $E_{\alpha,\delta}$ becomes: $E_{\alpha,\delta}(\lambda)=E_{\alpha,\delta}(\lambda)-E_{\alpha,\delta}(\lambda_0)$. Unlike the absolute photocentre coordinates $E_{\alpha,\delta}$ which gives barycentric information on the global stellar image, $\epsilon_{\alpha,\delta}$ gives the barycentric information depending on the baselines used. Thus, the (u,v) coverage should be distributed as regularly and densely as possible, especially for the fast rotators with their flattened shape.
Note that the polar line profile is less deep than the equatorial one, which plays an important role on the final $E_{\alpha,\delta}(\lambda)$ values.
As soon as we measure $\phidiff$ on two different baselines $\vec{B}_{\rm i}$ and $\vec{B}_{\rm j}$ ($\rm i \ne j$), it yields an interferometric vectorial photocentre displacement $\vec{\epsilon}(\lambda)$.
To deduce the equatorial and polar photocentre displacement $(\epsilon_{\rm eq},\epsilon_{\rm pol})$ from $(\epsilon_{\alpha},\epsilon_{\delta})$, we must rotate the coordinate system by the rotation axis position angle $\rm PA_{\rm rot}$ (idem for $(E_{\rm eq},E_{\rm pol})$ from $(E_{\alpha},E_{\delta})$):
\begin{subeqnarray}\label{epseqpol_eq}
& \epsilon_{\rm eq}=\epsilon_\alpha\cos \rm PA_{\rm rot}+\epsilon_\delta\sin \rm PA_{\rm rot},  \\
& \epsilon_{\rm pol}=-\epsilon_\alpha\sin \rm PA_{\rm rot}+\epsilon_\delta\cos \rm PA_{\rm rot},
\end{subeqnarray}
Which means that $\epsilon_{\rm eq}$ and $\epsilon_{\rm pol}$ are completely independent of $\rm PA_{\rm rot} $. Taking $\rm PA_{\rm rot}=0^\circ$, $\epsilon_\alpha$ represents the equatorial photocentre displacement ($\epsilon_\alpha=\epsilon_{\rm eq}$) and $\epsilon_\delta$ the polar one ($\epsilon_\delta=\epsilon_{\rm pol}$).

\subsection{Discussing the photocentre displacements}
\label{photocentres2}

\begin{figure*}
\centering
 \includegraphics[width=0.44\hsize,draft=false]{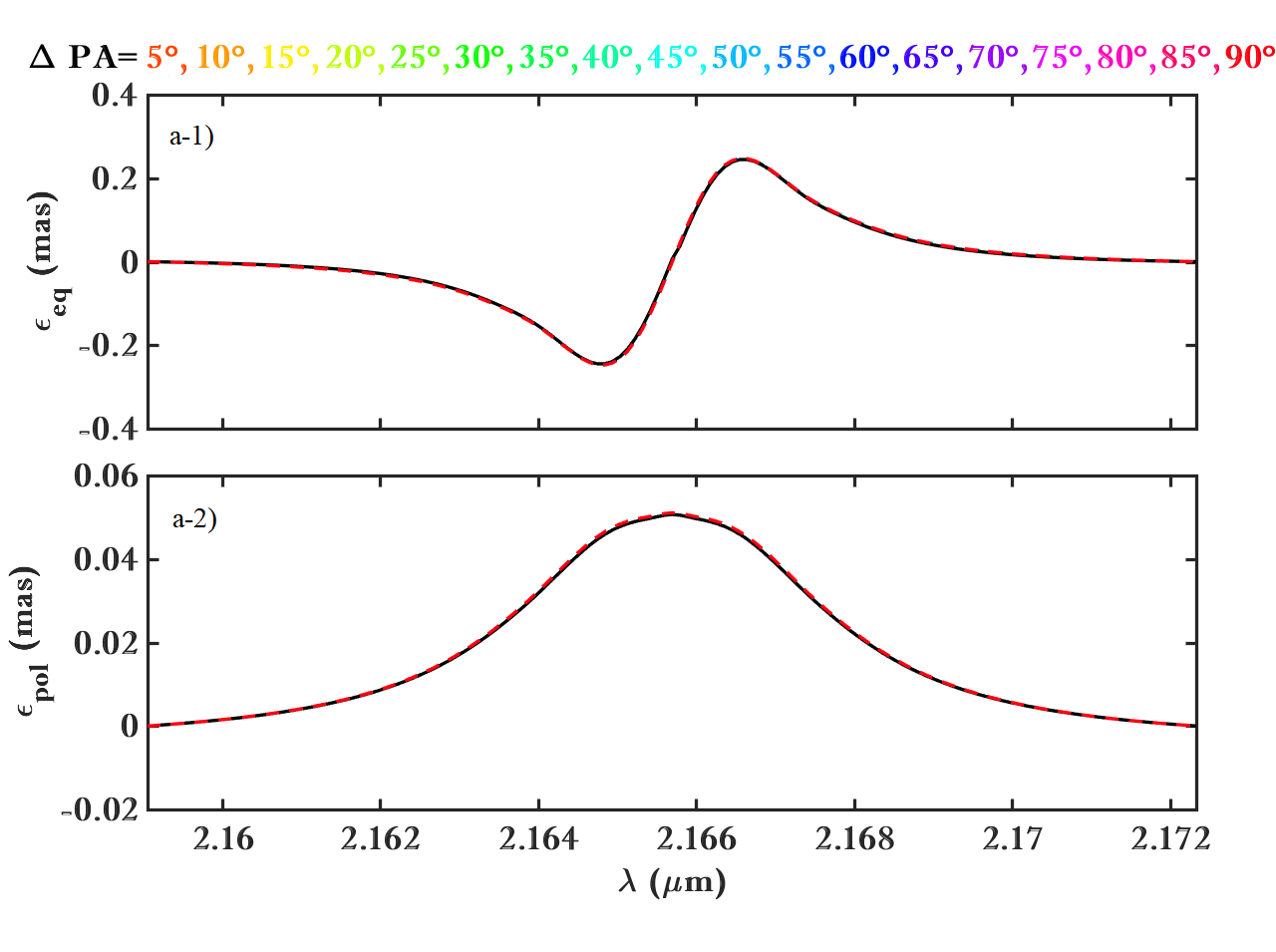}
 \includegraphics[width=0.44\hsize,draft=false]{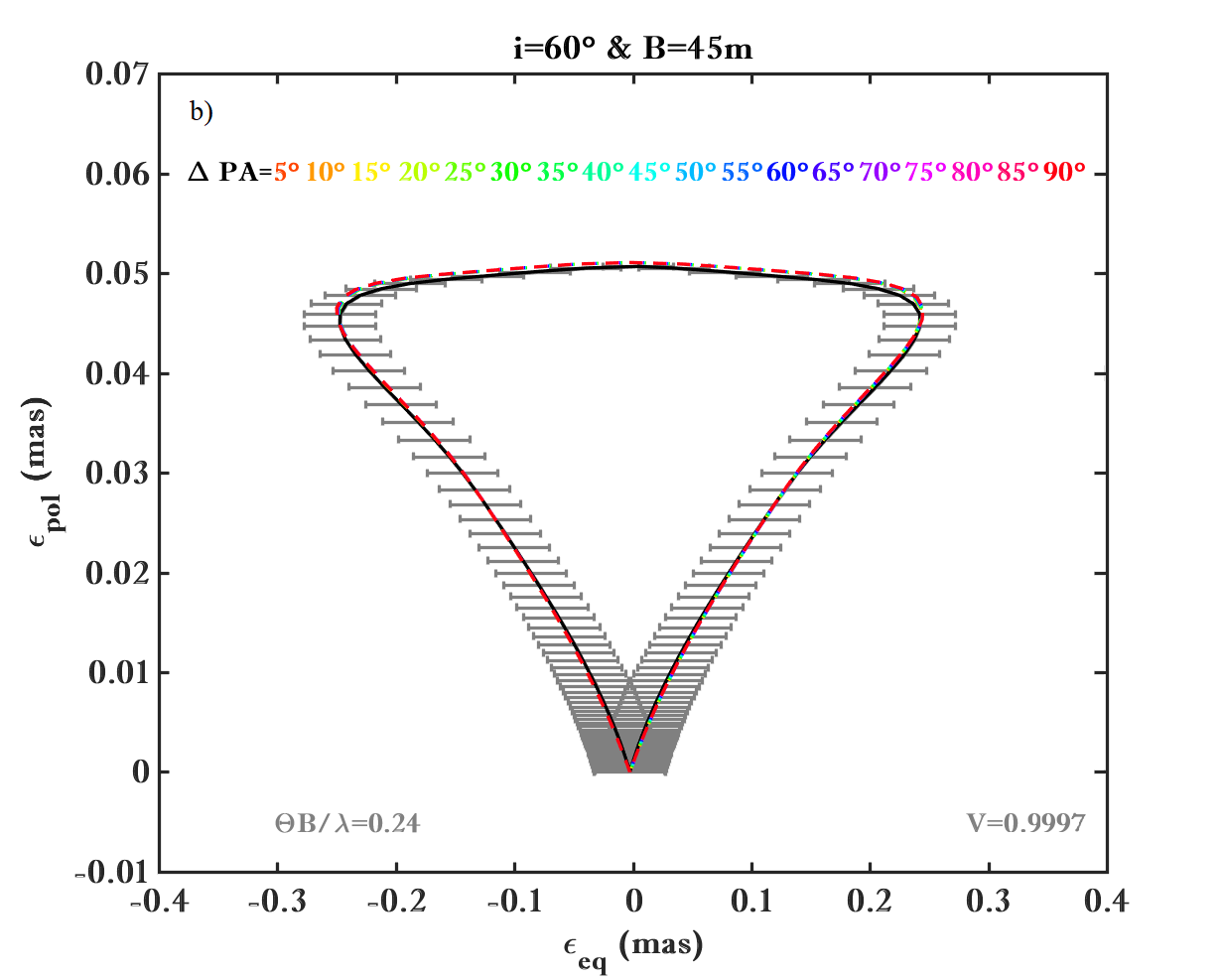}
 \includegraphics[width=0.44\hsize,draft=false]{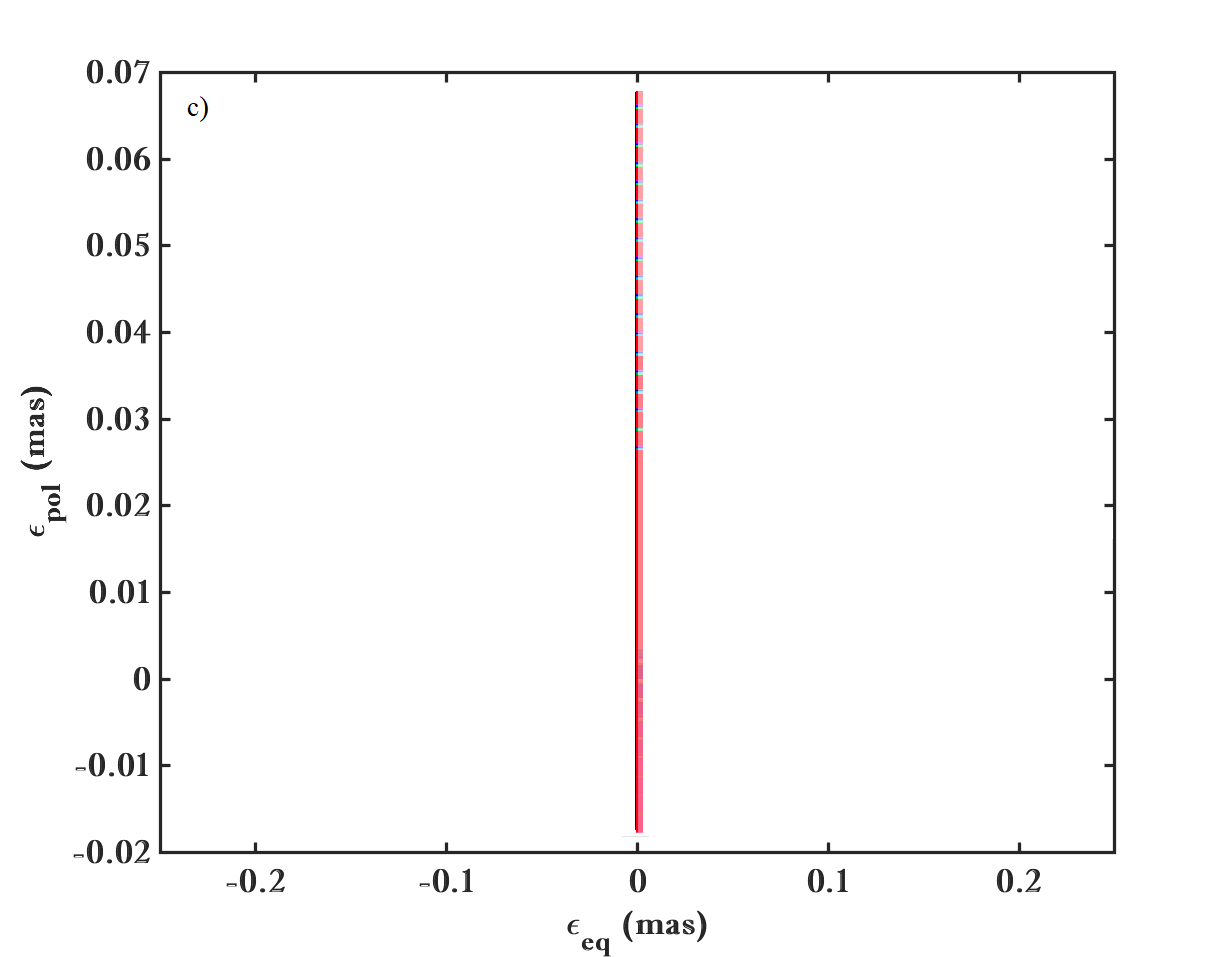}
 \includegraphics[width=0.44\hsize,draft=false]{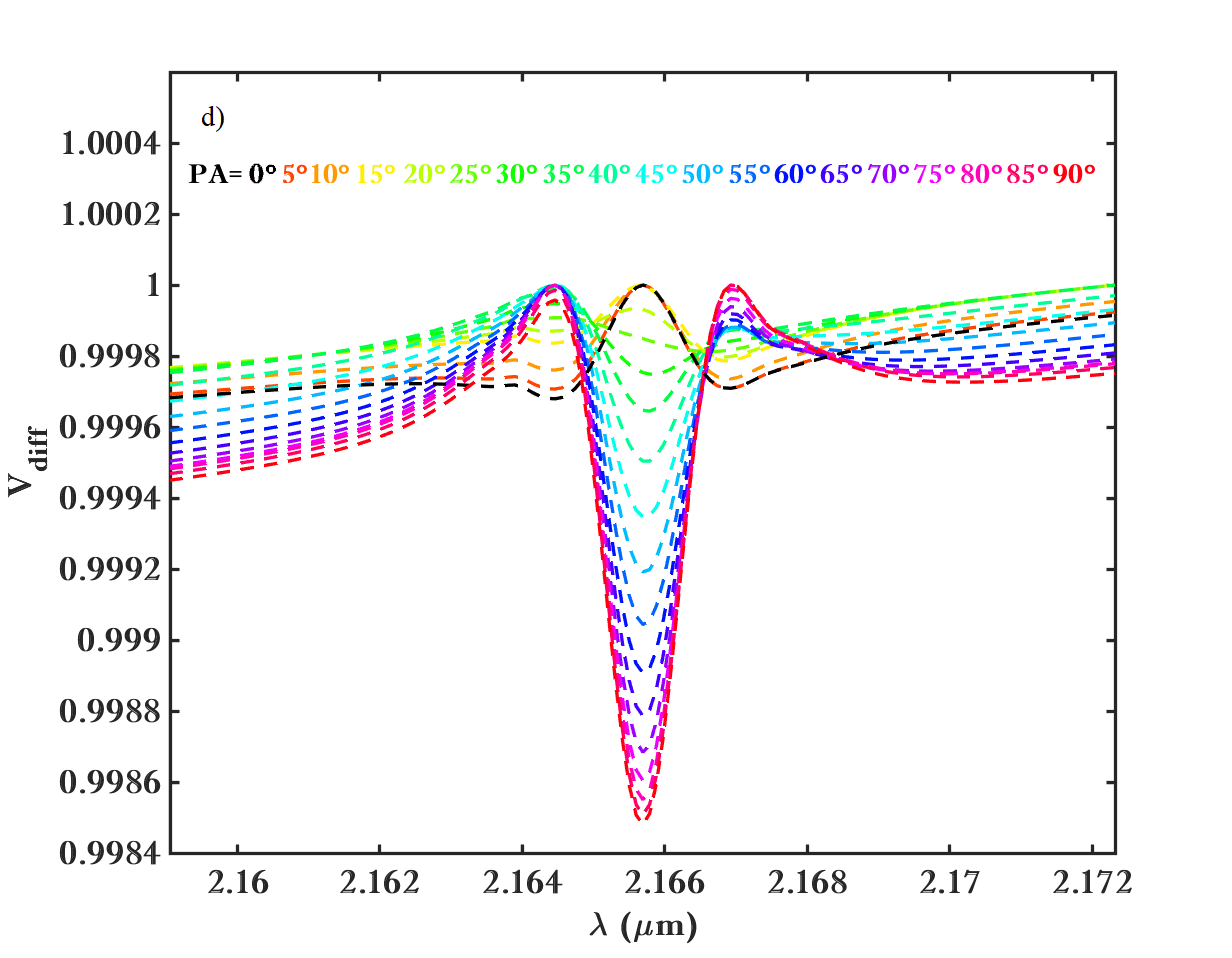}
 \caption{Absolute photocentres $E_{\rm eq,pol}$ (in black) and photocentres coming from the (u,v) coverage $\epsilon_{\rm eq,pol}$ (in colours), around the Br$\gamma$ line, for different position angle differences between two baselines ($\Delta \rm PA$, where each value corresponds to a specific colour on the top of each plot). \textbf{a)} $\epsilon_{\rm eq}$ and $\epsilon_{\rm pol}$ as a function of the wavelength. \textbf{b)} The vectorial photocentre displacement. The gray transparent box depicts the error of $30\,\mu as$ amplitude related to the measurement of $\epsilon_{\rm eq,pol}$. For aesthetic reasons we plot here only $\epsilon_{\rm eq}$ errors. \textbf{c)} Values of $\rm PA_{\rm rot}$, which are obtained by a simple linear fit of the vectorial photocentre displacements. \textbf{d)} Corresponding differential visibilities at each baseline projection angle ($\rm PA$).} \label{Phot_Ach0}
\end{figure*}

Figure \ref{Phot_Ach0} shows a comparison between the absolute photocentre displacements $E_{\rm eq,pol}$ (continuous black line) with the apparent interferometric photocentre displacements $\epsilon_{\rm eq,pol}$ (discontinuous coloured lines) for a projection angle $i=60^\circ$, for only two baselines ($B=45\,\rm m$), with $\Delta \rm PA$ ranging from $5^\circ$ to $90^\circ$ by steps of $5^\circ$. For simplification reasons, in this subsection we only use two baselines. Of course, the method can be used and extended to more baselines, thanks to Eq.~\ref{phot-tot_eq}, as is done in Sec.~\ref{AnnexC}, and as was done by \cite{2018MNRAS.480.1263H} before.

Fig.~\ref{Phot_Ach0} (\textit{a}) shows the wavelength dependency of $\epsilon_{\rm eq}$ $\epsilon_{\rm pol}$.
Fig.~\ref{Phot_Ach0} (\textit{b}) shows the vectorial photocentre displacement ($\epsilon_{\rm pol}=f(\epsilon_{\rm eq})$), which is in shape of an arrowhead (further discussion are done about that on Sec.~\ref{AnnexC} and Appendix \ref{AnnexA}). For readability reason, we only show the uncertainties of $\epsilon_{\rm eq}$, because the uncertainties of $\epsilon_{\rm pol}$ cover all the frame, where both uncertainties are equal to $30\,\mu as$.
The error on the photocentre displacements $(\sigma_{\epsilon}=\pm30\,\mu as)$ was deduced before by \cite{2018MNRAS.480.1263H}, from RMS of $\phidiff$ errors of AMBER, in the continuum, for three VLTI baselines with an average length of $\rm B\sim 75\,\rm m$. The value of $\pm30\,\mu as$ is considered as a good average error of both $\epsilon_{\rm pol}$ and $\epsilon_{\rm eq}$ over all differential VLTI instruments in the IR.
Fig.~\ref{Phot_Ach0} (\textit{c}) depicts deduced $\rm PA_{\rm rot}$ values, through a simple linear fit (with $90^\circ$ rotation) of the vectorial photocentre displacements \citep[further details in our previous paper;][]{2018MNRAS.480.1263H}, which was initially set at $\rm PA_{\rm rot}=0^\circ$ for simplicity. 
Fig.~\ref{Phot_Ach0} (\textit{c}) shows the different differential visibilities corresponding to each baseline projection angle ($\rm PA$). The longer baseline length (B) the lower is the visibility. Indeed, the polar visibility at $\rm PA=0^\circ$ is lower than the equatorial one at $\rm PA=90^\circ$.
As we can observe, independently of $\rm PA$ value, which affect single photocentre displacement projections, and the visibilities, the combination/averaging of photocentre displacement coming form couple(s) is completely independent form $\rm PA$. The different plots of $\epsilon_{\rm eq,pol}$ as a function of $\rm PA$ are all superimposed on top of each other, for unresolved objects.

As we know, Eq.~\ref{phot-phi_eq} is not at all satisfied for the resolved objects. In other words, the photocentre method could not be used then, where other terms of $\phidiff$ start to alter the interferometric photocentre displacement. Therefore, one of the main objectives of this study is to determine the resolution range in which the photocentre method remains applicable. Indeed, the equality $E_{\rm eq,pol}\approx\epsilon_{\rm eq,pol}$ depends mainly on the angular size of the observed object ($\Theta_{\rm eq,pol}$), and relatively to the angular resolution $\lambda/B$). Thereby, we use $\Gamma_\epsilon=\Theta\frac{B}{\lambda}$ as criterion for the limits of use of photocentre displacement in interferometry.

Thus, by varying $\Theta$, $B$, and $\lambda$, allows us to determine the limit values of $\Gamma_\epsilon$, for which the photocentre displacement method is applicable, i.e. when the equality $E_{\rm eq,pol}\approx\epsilon_{\rm eq,pol}$ is satisfied. We show, in Appendix \ref{AnnexA} a few examples related to the case of a star similar to Achernar \citep[$\Theta=2.38\, \rm mas$][]{2014A&A...569A..45H}, but with $\rm PA_{\rm rot}=90^\circ$, and gravity darkening coefficient $\beta=0.25$, observed in the K-band (at $\lambda=2.16570\,\mu\rm m$), and for several values of $B=15$, $30$, $45$, $60$, $75$, $90$ until $150\,\rm m$.
Also, to study the sensitivity of $\Gamma_\epsilon$ to the inclination angle, we set $i= 30^\circ$, $60^\circ$, and $90^\circ$. Therefore, and according to all our tests, which are represented by the sample of figures (\ref{A8a}-\ref{A8d}) gathered in Appendix \ref{AnnexA}, we observe that the photocentre displacement method works only for quasi unresolved stars, with $\Gamma_\epsilon\leq 0.32$, i.e. when the angular size of the observed object is approximately 30\% less than $\frac{\lambda}{B}$ ($\Theta \lesssim 0.3 \frac{\lambda}{B}$). We add Fig.~\ref{A8d} in Appendix \ref{AnnexA}, for a fixed baseline length ($B= 150\,\rm m$) and three different inclination angle values ($i= 30^\circ$, $60^\circ$, and $90^\circ$), in order to illustrate the case of a more resolved star.

From averaging couple photocentre projections (Fig.~\ref{Phot_Ach0} and Figs.~\ref{A8a}-\ref{A8d}), we observe that the more the star is edge-on ($i\sim 90^\circ$) the higher the equatorial photocentre amplitude (because of the higher $\vsini$), and the lower the polar photocentre amplitude (because of the gradient contrast variations along the polar axis in function of $i$). Formally, the more edge-on the star, the wider the arrowhead shape, and in contrary, the more pole-on the star ($i\sim 0^\circ$), the sharper the arrowhead shape. Moreover, because of the intensity gradient caused by the gravity darkening effect, we observe that the polar photocentre amplitude is inversely proportional to the inclination angle.

Therefore, we conclude that the vectorial photocentre displacement method can be used (i.e. when $E_{\rm eq,pol}\approx\epsilon_{\rm eq,pol}$) for $\Gamma_\epsilon\leq 0.32$, which corresponds to quasi unresolved objects, with a visibility $V \geq 0.99$. Beyond, i.e. $\Gamma_\epsilon > 0.32$, only $\phidiff$ data are usable. In other words the photocentre displacement method appears applicable only with unresolved stars. Of course, we can also use the ``classical'' differential phase method ($\phidiff$) in this case, but in terms of time computing, and when we have many baselines, on unresolved object, it is more efficient to use the vectorial photocentre displacements method. In contrary, for $\Gamma_\epsilon > 0.32$, only the ``classical'' $\phidiff$ method is useful.

The limit of $\Gamma_\epsilon$ depends of course on the $\vec{\epsilon}$ uncertainty that we have chosen ($30\,\mu as$). But in practice, with real data, we must not forget the noise effect (SNR). We could extend the application of the photocentre displacement method to marginally resolved stars $V\sim 0.8$ \citep[i.e. $\Gamma_{\epsilon}\sim 0.64$, as done before by][]{2018MNRAS.480.1263H} and a difference of the projection angle ($60^\circ\leq \Delta \rm PA\leq 90^\circ$), especially for edge-on star (with a very low amplitude of $\epsilon_{\rm pol}$).

\section{Dependence to model physical parameters}
\label{phot-sensitivity}

\begin{figure}
\centering
\includegraphics[width=1.\hsize,draft=false]{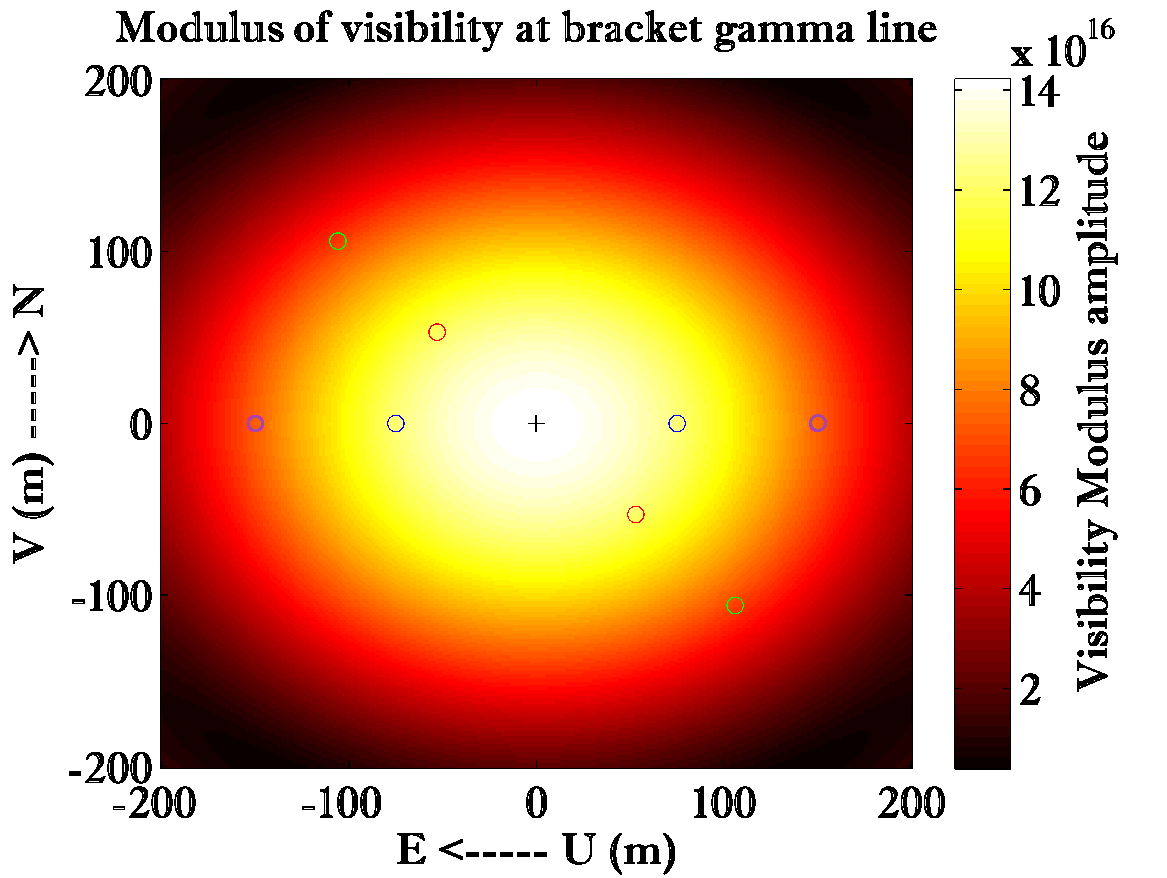}
\caption[module of visibility and (u,v) coverage of the reference model star] {(u,v) coverage and module of visibility of the reference model star similar to Achernar.} \label{uvCoverageAch_sim}
\end{figure}

As in \cite{2012A&A...545A.130D}, here we use four interferometric configurations with two baseline lengths $B_{\rm proj}=75\,\rm m$ and $150\,\rm m$ and two projection angles $\rm PA=45^\circ$ and $90^\circ$. In addition to the visibility modulus used for the reference modeling star, Fig.~\ref{uvCoverageAch_sim} shows the four $(u,v)$ coverage points used in our simulation.
The angular diameter of our first modeling reference star (Achernar-like) is of $2.38\pm0.10\,\rm mas$ \citep{2014A&A...569A..45H} and the angular resolution corresponding to our baselines $B_{\rm proj}=75\,\rm m$ ($\Gamma_\epsilon=0.4$) and $B_{\rm proj}=150\,\rm m$ ($\Gamma_\epsilon=0.8$) is of $5.96\rm\,mas$ and $2.98\rm\,mas$ respectively. 

In this study, we use 4 interferometric configurations (of 2 telescopes), as shown in Fig.~\ref{uvCoverageAch_sim}, which represents the visibility modulus used for our reference star (Achernar-like). Our objective is to study the dependence of the flux and photocentres displacement on relevant physical parameters of our model, namely: the stellar radius $R$, $\vsini$, inclination angle $i$, average effective temperature $\Tmean$, with and without limb-darkening, for different line profiles (Voigt, from Kurucz and Tlusty modeling), fixed and as function of the co-latitude $\theta$. The dependence of the gravity darkening coefficient $\beta$ will be studied in the Sec.~\ref{sensitiv_eps}.
Note that the West-to-East direction is taken from right to left in all plots of the current paper, including in the previous Fig.~\ref{Phot_Ach0}.
At this step, we have to remind that we could use a Voigt function as simple and rough line profile but, and as explained in Appendix.~\ref{AnnexAlpha}, we use the pseudo-Voigt function (of 6 parameters) only to get an analytic expression of the Kurucz/Tlusty line profiles for faster calculation purpose. So, we study in the present section the difference between a Voigt line profile and a pseudo-Voigt fit function of the line profiles from Kurucz/Tlusty model atmospheres.

The influence of each studied parameter on the simulated spectrum and perpendicular photocentre displacements ($\epsilon_\alpha$ and $\epsilon_\delta$) on the $Br_\gamma$ line are shown in Figs \ref{fig_model1} and \ref{fig_model2}, in Appendix \ref{AnnexB} below.
We note that, in our case where $\rm PA_{\rm rot}=0^\circ$, $(\epsilon_{\rm eq},\epsilon_{\rm pol})=(\epsilon_\alpha,\epsilon_\delta)$ and $(E_{\rm eq},E_{\rm pol})=(E_{x},E_{y})$.

For practical reasons, all the figures for this section are gathered in Appendix \ref{AnnexB}. Thereby, and as we can observe in Figs.~\ref{fig_model1} and \ref{fig_model2}, we can classify the effect of the stellar parameters on photocentre displacements (or on $\phidiff$), thanks to $\vec{\epsilon}$ differences of ($\Delta_{\epsilon}$) with respect to reference model, in three categories:

1a) The very sensitive parameters ($\Delta_{\epsilon}>25\%$), which are the equatorial radius $R_{\rm eq}$, $\rm PA_{\rm rot}$ and the line profile linked to the analytic/stellar atmosphere modeling (Voigt versus Kurucz).

1b) The moderately sensitive parameters ($10\%\leq\Delta_{\epsilon}\leq25\%$), which are $\vsini$, the inclination $i$, and the line profile linked to the analytic/stellar atmosphere modeling (Tlusty versus Kurucz, and the line profile linked to the latitudinal versus the regular aspect).

1c) And finally the less sensitive parameters ($\Delta_{\epsilon}<10\%$), which are the mean effective temperature $\Tmean$, $\beta$,the both darkening (limb and gravitational). 

The spectrum, on the other hand, is (the same, thanks to $\Delta_{s}$ with respect to reference model): 

2a) More sensitive (with $\Delta_{s}\geq5\%$) to the line profile linked to the analytic/stellar atmosphere modeling aspect; e.g. Voigt-Kurucz.  

2b) Moderately sensitive (with 1\%$\leq\Delta_{s}<5\%$) to other kind of line profiles (Tlusty-Kurucz and linked to the latitudinal/regular aspect), $\vsini$, $i$.

2c) And finally, less sensitive (with $\Delta_{s}<1\%$) to $R_{\rm eq}$, $\Tmean$, $\rm PA_{\rm rot}$.and to the limb-darkening. 

It should be noted that the study of the influence of the line profile on the interferometric measurements allows us to demonstrate the important impact of this parameter. Indeed we demonstrate the impossibility of using a simple analytical profile in this kind of simulation (e.g. a simple and rough Voigt line profile). The comparison of SCIROCCO code with the photocentre displacements observations via a particular adjustment method (explained in the section below) to four free parameters ($R_{\rm eq}$, $V_{\rm eq}$, $i$ and $\rm PA_{\rm rot}$), reveal a important dependence of the $R_{\rm eq}$ and $V_{\rm eq}$ dynamical parameters, which are also strongly correlated and related in the equation of degree of sphericity $D$ \citep{2002A&A...393..345D}:
\begin{equation}\label{deg-spher_eq}
 D=\frac{R_{\rm pol}}{R_{\rm eq}}=\left(1+\frac{V^2_{\rm eq}R_{\rm eq}}{2GM}\right)^{-1}\approx 1-\frac{V^2_{\rm eq}R_{\rm pol}}{2GM},
\end{equation}

where $G$ the gravitation Cavendish-Newton constant and $M$ the mass of the star. The parameter $V_{\rm eq}$ is highly sensitive to the type of line profile (about 13\%) of average difference between the Voigt line profile and the Kurucz/Synspec profile, while the geometrical parameters $R_{\rm eq}$, $i$ and $\rm PA_{\rm rot}$ remain less sensitive to these profiles.

\section{Application to Regulus}
\label{AnnexC}
In this section, we discuss in more details our methodology of the photocentre displacements and its application to real spectro-interferometric data. 
In a previous paper about the fast rotator Regulus observed with VLTI/AMBER, \cite{2018MNRAS.480.1263H} deduced the best fitting parameters, namely: the equatorial radius $R_{\rm eq}=4.16 \pm 0.24\,R_\odot$, the equatorial rotation velocity $V_{\rm eq}=350 \pm 22\,\kms$, the rotation-axis inclination angle $i=86.4 \pm 6.3^\circ$, and the rotation-axis position angle $\rm PA_{\rm rot}=251 \pm 2^\circ$, from differential phase data ($\phidiff$), using a $\chi^2$ minimization and the Markov Chain Monte Carlo method (MCMC) method \citep{s11222-006-9438-0}. \cite{2018MNRAS.480.1263H} showed the final results through corresponding photocentre displacements $(\epsilon_{\rm eq},\epsilon_{\rm pol})$. They also noticed a slight dissymmetry on the observed photocentre displacements, between the red and the blue wings. This dissymmetry was caused by calibration issue in wavelength of a differential phase measurement over the six ones used. By fixing this issue, and enhancing the wavelength calibration for the spectrum as well, we obtained symmetric data and centered around the Br$\gamma$ line, as shown in in Fig.~\ref{A1} below.
\begin{figure*}
\centering
\includegraphics[width=0.55\hsize,draft=false]{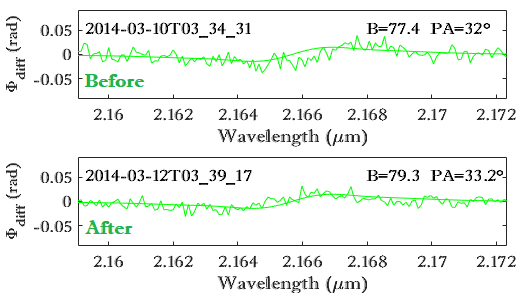}
\includegraphics[width=0.44\hsize,draft=false]{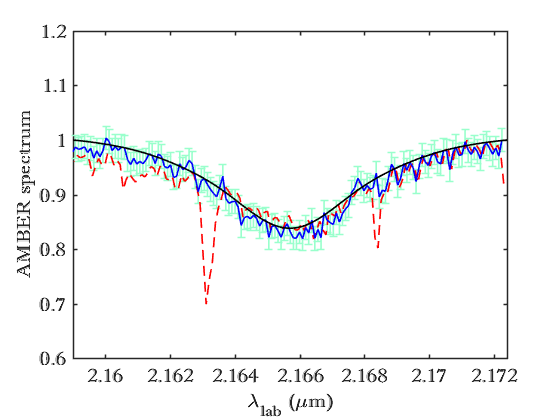}
\caption{\textbf{Left:} Effect of the wavelength calibration issue on the differential phase. Before the correction (top) and after the correction (bottom). The smooth thin green curve superimposed on the observation show the best $\phidiff$ model at this (u,v) coordinate. \textbf{Right:} AMBER spectrum of Regulus in the Br$\gamma$ line. The dashed thick red curve is the raw spectrum showing two telluric lines. The full thick blue curve is the Regulus spectrum, with its error bars in green. The thin dark line represents our best model.}
\label{A1}
\end{figure*}

\begin{figure*}
\centering
\includegraphics[width=0.48\hsize,draft=false]{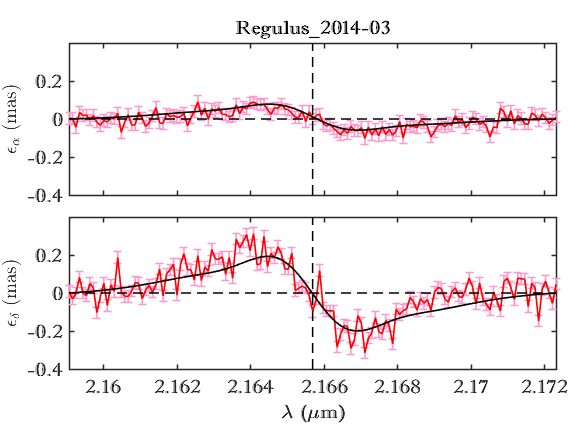}
\includegraphics[width=0.48\hsize,draft=false]{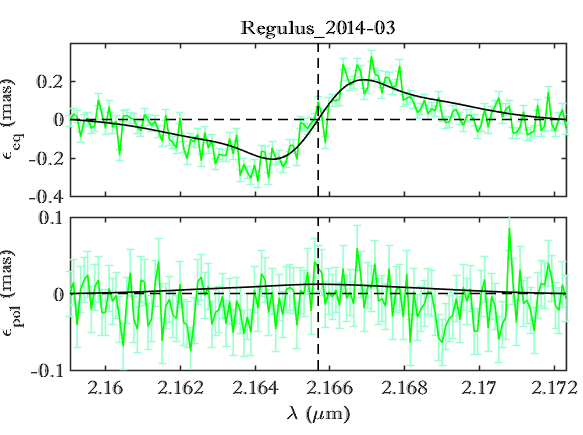}
\caption{\textbf{Left:} The perpendicular right ascension and declination photocentre displacements $(\epsilon_{\alpha},\epsilon_{\delta})$ as red thick curves for the observed data. The smooth thin black curves superimposed on the observations show the best-fitting $\phidiff$ (new MCMC results below). The two perpendicular dashed lines represent the zero-point for the photocentre displacement axis and the central wavelength ($\lambda$ = 2165.7\,\rm nm) of the Br$_\gamma$ line. \textbf{Right:} The equivalent equatorial–polar photocentre displacements $(\epsilon_{\rm eq},\epsilon_{\rm pol})$.}
\label{A3}
\end{figure*}

\begin{figure*}
\centering
\includegraphics[width=1.1\hsize,draft=false]{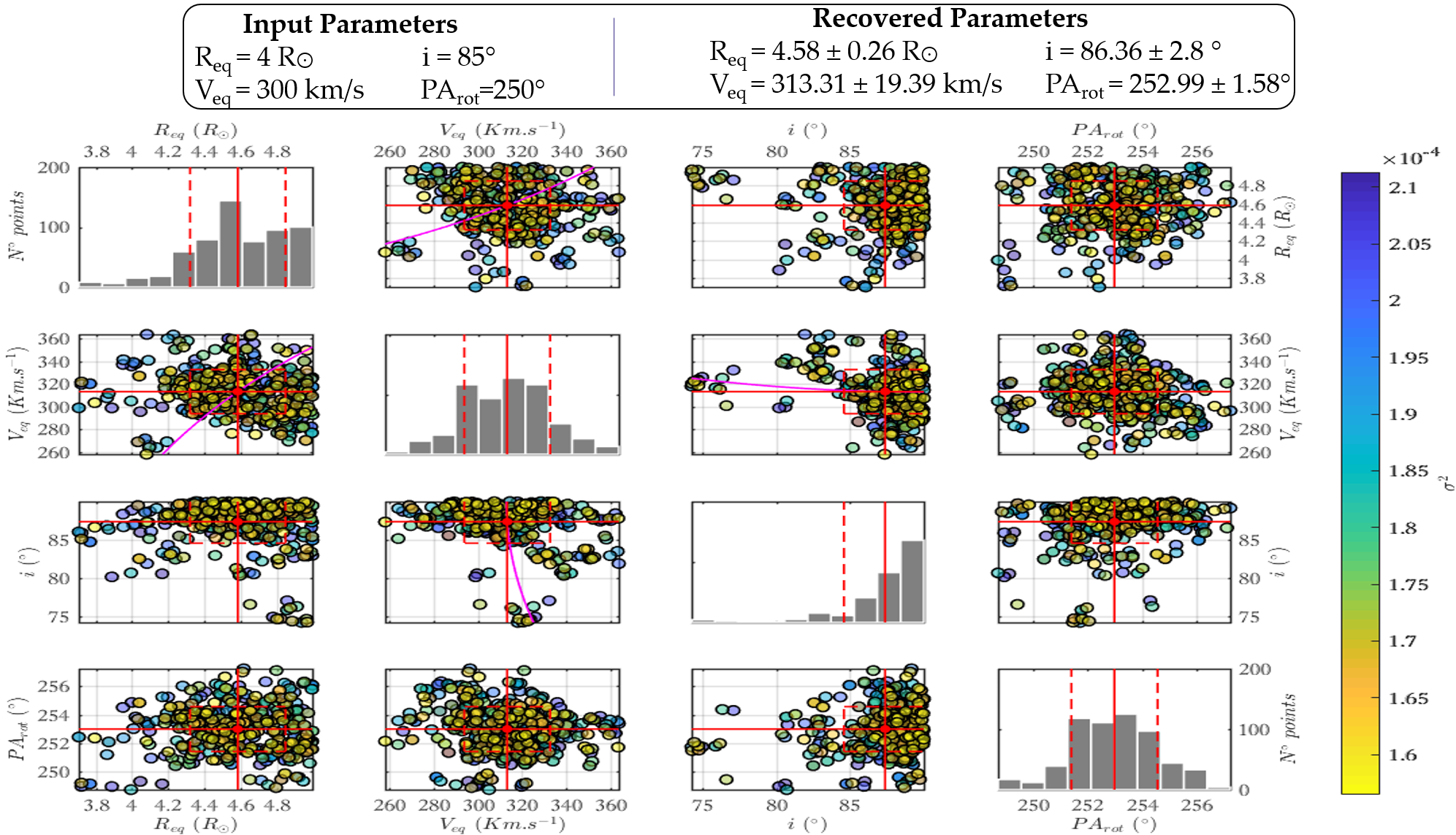}
\caption{MCMC covariance matrix distribution results for the four free parameters ($R_{\rm eq}$, $V_{\rm eq}$, $i$ and $\rm PA_{\rm rot}$) of Regulus, using six differential phases and spectrum data. The red point and line show the best recovered parameters, the average of the last MCMC run. The scatter plots show the projected two-dimensional distributions of the projected covariance matrix (coloured points) by pair of parameters. The colour bar represents the distribution of the points around the average, following the variance $\sigma^2$. The histograms show the projected one-dimensional distributions, with solid red lines representing the best recovered parameters and dashed red line the uncertainties.}
\label{A5}
\end{figure*}

Also, \cite{2018MNRAS.480.1263H} did not take into account the slight slope seen with the observed $(\epsilon_{\rm eq},\epsilon_{\rm pol})$, which is caused by an instrumental bias related to the spectrograph of AMBER, and that we treat and fix in the current section. Thereby, we fix this bias by a simple linear fit on the continuum, which we remove at the end, as shown in Fig.~\ref{A2} below.
\begin{figure*}
\centering
\includegraphics[width=0.48\hsize,draft=false]{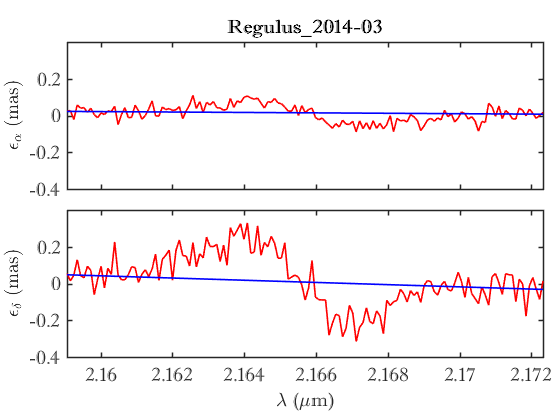}
\includegraphics[width=0.48\hsize,draft=false]{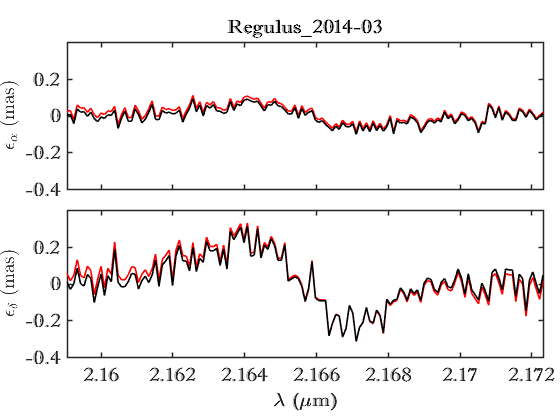}
\caption{\textbf{Left:} Observed perpendicular right ascension and declination photocentre displacements $(\epsilon_{\alpha},\epsilon_{\delta})$ in red thick curves, and their simple linear fits at the continuum in blue lines. \textbf{Right:} The same as in left plots, but in addition, corrected $(\epsilon_{\alpha},\epsilon_{\delta})$ in superimposed black lines (after removing the left blue lines deduced by a simple linear fit on the continuum).}
\label{A2}
\end{figure*}
Thereby, Fig.~\ref{A3} shows the final perpendicular right ascension and declination photocentre displacements $(\epsilon_{\alpha},\epsilon_{\delta})$, and their equivalent equatorial–polar photocentre displacements $(\epsilon_{\rm eq},\epsilon_{\rm pol})$, deduced by rotation thanks to the rotation-axis position angle \citep[$\rm PA_{\rm rot}$;][]{2018MNRAS.480.1263H}. The RMS error per spectral channel has been measured in the continuum outside the spectral line and found to be $\sim30 \mu as$ on any projection.
Once these corrections were made, the plot of our best modeling of equatorial–polar photocentre displacements ($\epsilon_{\rm pol}=f(\epsilon_{\rm eq})$) is an arrowhead-shaped curve, as shown in the Fig.~\ref{A4} below, with their new MCMC results. This arrowhead shape characterizes fast rotators. For edge-on rotators (i.e. $i\sim 90^\circ$), the faster the stellar rotation, the wider the arrowhead shape (as shown previously in Sec.~\ref{sensitiv_eps}).
\begin{figure*}
\centering
\includegraphics[width=0.48\hsize,draft=false]{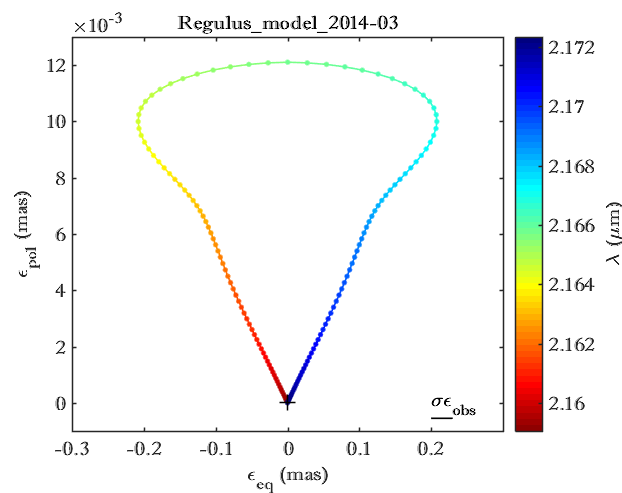}
\caption{Best modeling of vectorial photocentre displacement $(\epsilon_{\rm pol}=f(\epsilon_{\rm eq})$, which is in shape of arrowhead.}
\label{A4}
\end{figure*}
Fig.~\ref{A5} shows the covariance matrix of pairs of parameters, with their histogram, that were obtained by the MCMC method, with these new corrected photocentre displacements. We used exactly the same algorithm, configuration and starting input parameters as in \cite{2018MNRAS.480.1263H}.
We found the following correlation coefficients (in descending order) : $\rho(i,\rm PA_{\rm rot}) = 0.1765$, $\rho(R_{\rm eq},\rm PA_{\rm rot}) = 0.0128$, $\rho(R_{\rm eq},i) = -0.0473$, $\rho(V_{\rm eq},\rm PA_{\rm rot}) = -0.1486$, $\rho(R_{\rm eq},V_{\rm eq}) = -0.1837$ and $\rho(V_{\rm eq},i) = -0.2349$. When $\rho(A,B) > 0$, the correlation is proportional ($A$ and $B$ increase or decrease together), and when $\rho(A,B) < 0$ the correlation is inversely proportional ($A$ increases when $B$ decreases and vice versa), e.g. the case of $R_{\rm eq}$ and $V_{\rm eq}$, because of angular momentum conservation. Note also that the correlation coefficients of all our free MCMC parameters are symmetric (i.e. $\rho(A, B) = \rho(B, A)$).

These MCMC results are consistent with those given in a previous paper \citep[see MCMC section of][]{2018MNRAS.480.1263H}. The equatorial velocity $V_{\rm eq}$ is more consistent with previously published results, and $R_{\rm eq}$ is slightly higher but both, within their uncertainties, are closer to previous results found in the literature, compared to \cite{2018MNRAS.480.1263H}, especially for $V_{\rm eq}$.

\begin{figure*}
\centering
\includegraphics[width=1.1\hsize,draft=false]{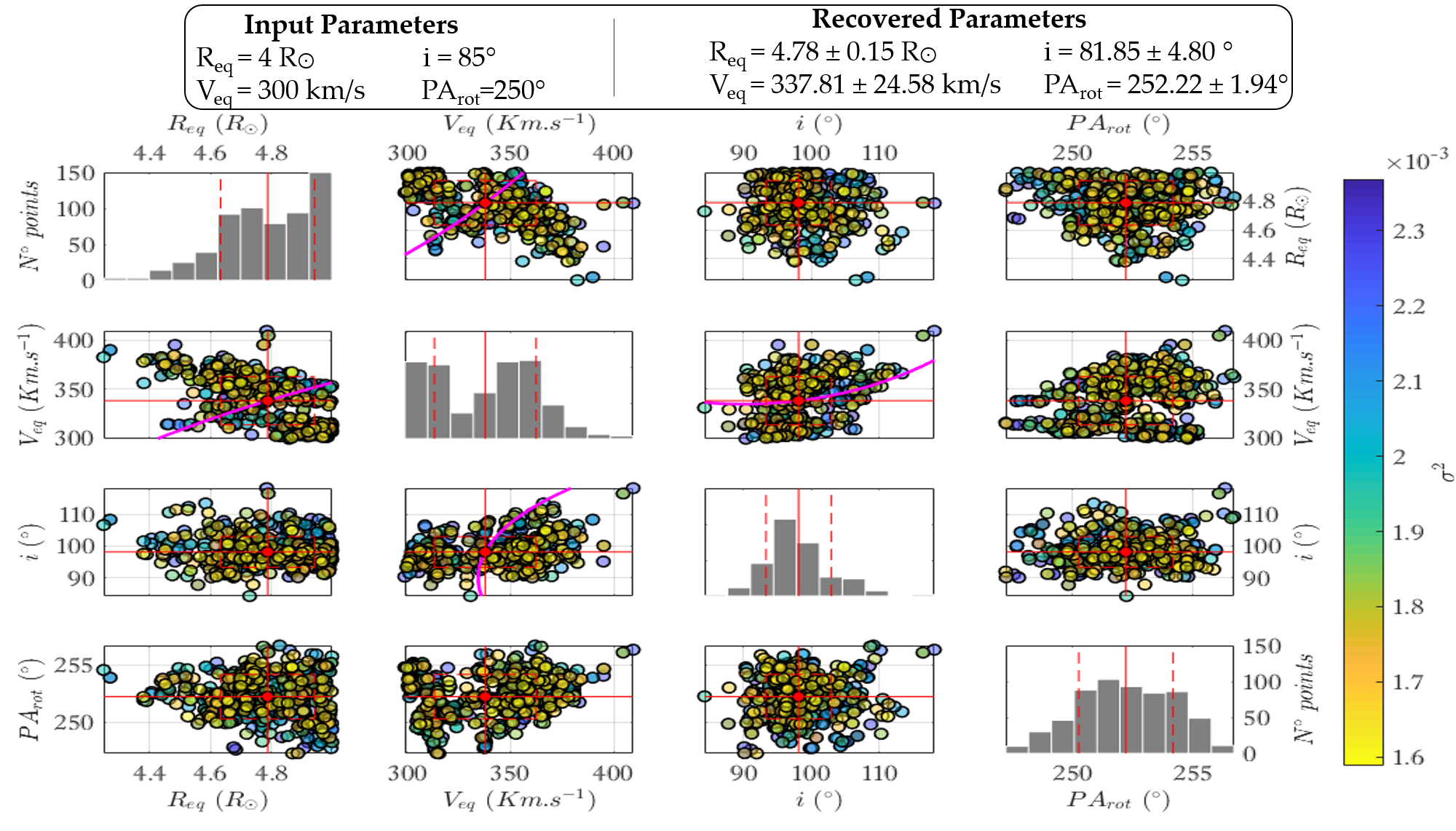}
\caption{Same as Fig.~\ref{A4}, but using two photo-centre displacements $(\epsilon_{\alpha},\epsilon_{\delta})$ and the spectrum data.}
\label{A6}
\end{figure*}

\begin{figure*}
\centering
\includegraphics[width=0.44\hsize,draft=false]{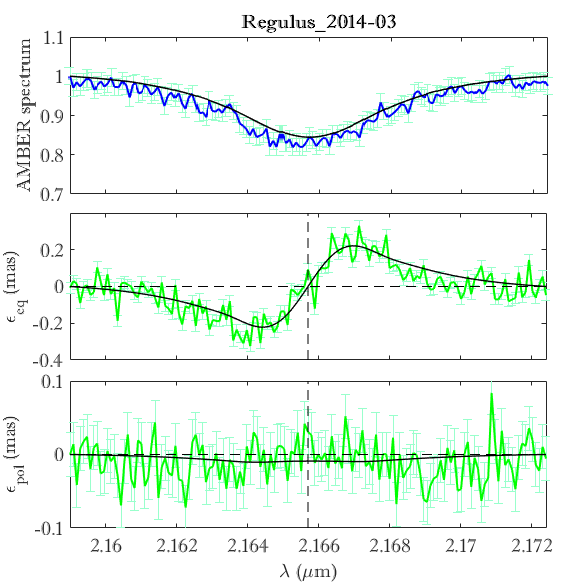}
\includegraphics[width=0.54\hsize,draft=false]{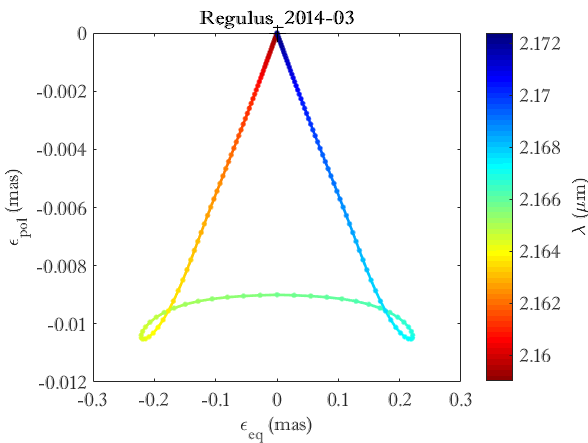}
\caption{Comparison between observations and the best MCMC minimization parameters model, through the two photocentre displacements $(\epsilon_{\alpha},\epsilon_{\delta})$ and spectrum, \textbf{Top:} over the AMBER spectrum of Regulus in the Br$\gamma$ line. \textbf{Bottom:} The equatorial-polar photocentre displacements $(\epsilon_{\rm eq},\epsilon_{\rm pol})$. \textbf{Right:} Best modeling of vectorial photocentre displacement $(\epsilon_{\rm pol}=f(\epsilon_{\rm eq})$.}
\label{A7}
\end{figure*}

These results were obtained by using differential phases ($\phidiff$) and spectrum data. But dealing with the two photocentre displacements $(\epsilon_{\alpha},\epsilon_{\delta})$ instead of the 6 values of $\phidiff$, in addition of the spectrum values, we obtain the results which are summarized in Fig.~\ref{A6}.
The correlation coefficient are: $\rho(V_{\rm eq},i) = 0.3594$, $\rho(V_{\rm eq},\rm PA_{\rm rot}) = 0.2598$, $\rho(i,\rm PA_{\rm rot}) = 0.1083$, $\rho(R_{\rm eq},i) = -0.0567$, $\rho(R_{\rm eq},\rm PA_{\rm rot}) = -0.1653$, and $\rho(R_{\rm eq},V_{\rm eq}) = -0.6070$. Relatively to the previous results, these new results are similar within their uncertainties, including for the inclination angle $i=81.8\pm4.8^\circ$ (visible south pole), instead of $86.4\pm2.8^\circ$ (visible north pole). The $i$-value of $81.8\pm4.8^\circ$ (south pole-on), is clearly consistent with what we find in the literature \citep[][and references therein]{2018MNRAS.480.1263H}. This slight difference of visible south/north poles could be explained by the fact that Regulus is marginally resolved (V$\sim$0.8 to 0.9), and the photocentre displacements method works only for unresolved cases (V$\sim$1). Note also that $\rm PA_{\rm rot}$-values are the same (i.e. $\sim 252^\circ$) in both studies.
Figure \ref{A7} depicts a comparison between observations and the best MCMC minimization parameters model, through the two photocentre displacements $(\epsilon_{\alpha},\epsilon_{\delta})$ and spectrum, over the AMBER spectrum of Regulus in the Br$\gamma$ line, the equatorial-polar photocentre displacements $(\epsilon_{\rm eq},\epsilon_{\rm pol})$, and $(\epsilon_{\rm pol}=f(\epsilon_{\rm eq})$ respectively. Unlike the Fig.~\ref{A4}, where the arrowhead points down, because of visible north pole, the arrowhead of Fig.~\ref{A7} points up (visible south pole). For this reason, the $\epsilon_{\rm pol}$ displacement have opposite signs in Fig.~\ref{A4} (positive) and in Fig.~\ref{A7} (negative). Because of the von Zeipel effect, the polar photocentre displacement $(\epsilon_{\rm pol})$ is always higher at the center of the line. For visible north pole, and as we can observe in Fig.~\ref{A3}, $(\epsilon_{\rm pol})$ increases from the red wing continuum until the line's center before decreasing again to the blue wing continuum. But in the case of visible south pole, as we can observe in Fig.~\ref{A7}, $(\epsilon_{\rm pol})$ decreases from the red wing continuum before increasing at the line's center, then it decreases before it increases again to the blue wing continuum, which explains the loops that can be seen on the sides of the arrowhead pointing up  of Fig.~\ref{A7}, and their absences on the arrowhead pointing down of Fig.~\ref{A4}. These ``loops'' disappear at an inclination angle $i\geq 10^\circ$ (visible south pole), when the intensity/temperature gradient is more pronounced on $\epsilon_{\rm pol}$, as we can observe in Fig.~\ref{horseshoes}.

In the next section we study the sensitivity of the vectorial photocentre displacement to the main key parameters of the fast rotators, namely: $R_{\rm eq}$, $V_{\rm eq}$, $i$ and $\rm PA_{\rm rot}$ for the case of Regulus, where the line profile varies along the co-latitude $\theta$ according to the $\beta$-value.

\section{Sensitivity to $R_{\rm eq}$, $V_{\rm eq}$, $i$, $\rm PA_{\rm rot}$, and $\beta$}
\label{sensitiv_eps}

\begin{figure*}
\centering
\includegraphics[width=0.48\hsize,draft=false]{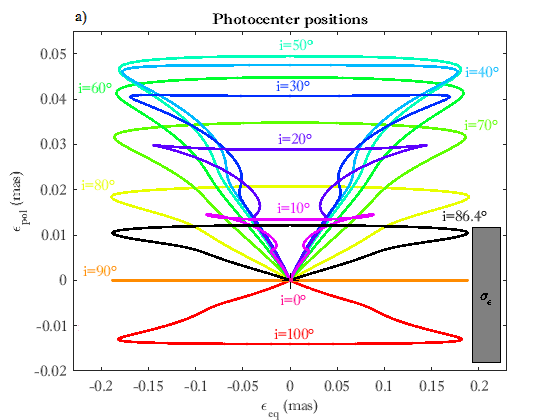}
\includegraphics[width=0.48\hsize,draft=false]{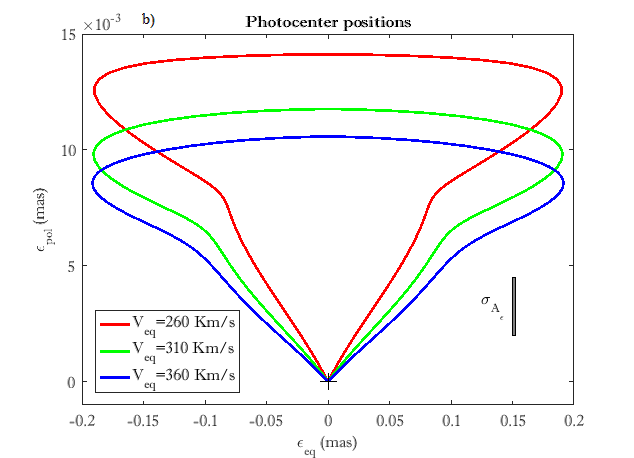}
\includegraphics[width=0.48\hsize,draft=false]{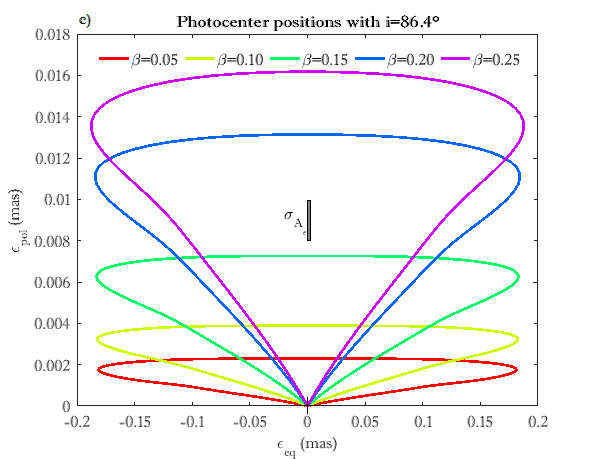}
\includegraphics[width=0.48\hsize,draft=false]{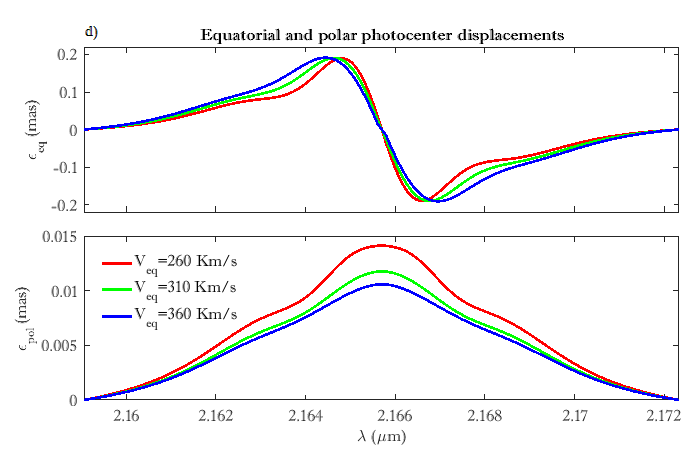}
\caption{\textbf{Top-Left (Fig.~13a):} Sensitivity of the photocentre displacement to the inclination $i$ with an enhancement of the polar scale, for the parameters of our best model \citep[$R_{\rm eq}=4.16\,R_\odot$, $V_{\rm eq}=350\,\kms$, $i=86.4^\circ$, and $PA_{\rm rot}=251^\circ$; see Sec.5 of our previous paper][]{2018MNRAS.480.1263H}, which are close of those used by \protect\cite{2011ApJ...732...68C}, including $\beta\approx 0.16$. The error box with $30\,\mu \rm as$ edges represents the error of each spectral measure of $\epsilon$. 
\textbf{Top-right (Fig.~13b):} Sensitivity of the photocentre displacement to the equatorial velocity $V_{\rm eq}$. Same plots as in Fig.~13a for $i=86.4^\circ$ and various values of $V_{\rm eq}$. The error box with $2.5\,\mu \rm as$ represents the minimum error on the mean amplitude of $\epsilon_{\rm eq}$ or $\epsilon_{\rm pol}$ estimated from all channels in the line. 
\textbf{Bottom-Left (Fig.~13c):} Sensitivity of the photocentre displacement to the gravitational darkening parameters $\beta$. Same plot as in Fig.~13a with $i=86.4^\circ$, $V_{\rm eq}=350\,\kms$ and $\beta$ ranging from 0.05 to 0.25. 
\textbf{Bottom-Right (Fig.~13d):} The equatorial and polar photocentre displacements for different $V_{\rm eq}$ with the same parameters than in Fig.~8b.}
\label{horseshoes}
\end{figure*}

In this section we use the SCIROCCO code to model the almost edge-on fast rotator Regulus \citep[using the best-fitting parameters given by][with $i=86.4^\circ$]{2018MNRAS.480.1263H}, in order to study the sensitivity of the differential photocentre to $R_{\rm eq}$, $V_{\rm eq}$, $i$, $\rm PA_{\rm rot}$, and especially to the gravitational darkening parameter $\beta$. Indeed, an edge-on fast rotator allows a better study of the gravity darkening parameterr on the photocentre displacements.
In Fig.~\ref{horseshoes} we illustrate the effect of our key parameters ($R_{\rm eq}$, $V_{\rm eq}$, $i$, $\rm PA_{\rm rot}$, and $\beta$) on the photocentre displacement, for Regulus with a line profile which varies along the co-latitude $\theta$ according to $\beta$-value \citep[as we did on][]{2018MNRAS.480.1263H}.
An important and complementary section of discussion (Sec.~\ref{AnnexC}) allows a better comprehension of the results shown below. 

In the panels \ref{horseshoes}a, \ref{horseshoes}b and \ref{horseshoes}c we see the effect of the inclination, the equatorial velocity and the gravitational darkening on a plot of the photocentre location as a function of wavelength. The vertical axis is in the polar direction and the horizontal one in the equatorial direction.
It is obvious that all these curves are symmetric with respect to the polar axis. This confirms that the position angle of the rotation axis can be measured independently from all other parameters, for example by a linear fit through all photocentre points as we did in the previous paper \citep{2018MNRAS.480.1263H}.

Note that all photocentre displacement are proportional to the stellar angular radius. However, panels \ref{horseshoes}a and \ref{horseshoes}b show that the total amplitude of $\epsilon_{\rm eq}$ (the width of the curves) is independent from $\vsini$ within a large range of velocities and inclinations. Panel \ref{horseshoes}c also shows that the sensitivity of this amplitude $A_{\epsilon_{\rm eq}}$ to the gravitational darkening parameter $\beta$ is not negligible. We should therefore have access to an accurate measure of $R_{\rm eq}$ from the equivalent width of $\epsilon_{\rm eq}(\lambda)$. Here the total amplitude $A_{\epsilon_{\rm eq}}\approx 400\,\mu as$ and the average width of the figure \ref{horseshoes}a can be roughly estimated to be $A_{\epsilon_{\rm eq}}/2 \approx 400 \mu as$. If we combine all 141 measures in the line with independent errors of $30\,\mu \rm as$ we get a global error on the half amplitude of $2.5\,\mu \rm as$ and a relative error of 1.25\% on the angular radius, with little influence from the other parameters discussed here.

Panels \ref{horseshoes}b and \ref{horseshoes}a show that the velocity and inclination together, i.e. $\vsini$ mainly changes the polar amplitude $A_{\epsilon_{\rm pol}}$, which increases with $|90^\circ-i|$ for $45^\circ$ as maximum in panel \ref{horseshoes}a. Around $350\,\kms$, the variation $A_{\epsilon_{\rm pol}}/\vsini\approx 0.05\,\mu\rm as/\kms$ yields a very large error on the radial velocity $\sigma_{\vsini}\approx 50\,\kms$ in panel \ref{horseshoes}b. The  photocentre measurements that we use here provide a poor constraint on the equatorial velocity.
However, for a fixed $\vsini$ constrained by the line profile broadening, $A_{\epsilon_{\rm pol}}$ varies substantially with the inclination $i$. Panel \ref{horseshoes}a shows that for $80^\circ < i < 90^\circ$, the average amplitude $A_{\epsilon_{\rm pol}}$ changes by typically 1 $\mu \rm as/^\circ$ . This leads to an accuracy $\sigma_i \approx 2.5^\circ$ in the inclination, if all other parameters are well constrained.

\section{Conclusions and discussion}
\label{conclusions}

We have established in this paper the general formalism of the photocentre displacements and studied the impact of some relevant physical parameters for rapid rotators, using the semi-analytical model (SCIROCCO).

We show that even with a poor $(u,v)$ coverage \citep[e.g. in][and in Sec.~\ref{AnnexC}]{2018MNRAS.480.1263H}, our method allows to obtain important information from photocentre displacements $\epsilon_{\alpha,\delta}$. Indeed, two different baselines are sufficient for the use of the vectorial photocentre displacement method.
On fully unresolved targets, $\phidiff$ reduces to the measurement of the vectorial photocentre displacement $\vec{\epsilon}$, that is accessible as soon as we have two different baselines, preferably nearly orthogonal for SNR optimization.

We classify in three categories the stellar parameters impacting the photocentres displacements (Fig.~\ref{fig_model1} and ~\ref{fig_model2}). The very sensitive parameters ($\Delta_{\epsilon}>25\%$) such the equatorial radius $R_{\rm eq}$, $\rm PA_{\rm rot}$ and the line profile linked to the analytic/stellar atmosphere modeling (Voigt versus Kurucz).
The moderately sensitive ones ($10\%\leq\Delta_{\epsilon}\leq25\%$) as the $\vsini$, the inclination $i$, and the line profile linked to the analytic/stellar atmosphere modeling (Tlusty versus Kurucz, and the line profile linked to the latitudinal versus the regular aspect). And finally the less sensitive parameters ($\Delta_{\epsilon}<10\%$) which are the mean effective temperature $\Tmean$, $\beta$,the both darkening (limb and gravitational). 
The spectrum, on the other hand, is more sensitive (with $\Delta_{s}\geq5\%$) on the line profile (concerning also the analytic/stellar atmosphere modeling aspect; Voigt-Kurucz), which was expected because the line profile approximation affects the spectrum. However, the spectrum is less sensitive (with $\Delta_{s}<5\%$) to other kind of line profiles, and to the darkenings, $R_{\rm eq}$, $\vsini$ ,$i$, $\rm PA_{\rm rot}$, $\Tmean$ and $\beta$.

The photocentre displacement is a powerful tool for studying the poorly resolved critical/fast and slow rotators. Indeed, in the case of critical/fast rotators, the approach that we propose (for poorly/marginally‐resolved stars) allows to simultaneously constrain the rotation‐axis position angle $\rm PA_{\rm rot}$, estimate the range of the rotation‐axis inclination angle $i$, and under certain conditions the gravity‐darkening coefficient $\beta$ \citep{2018MNRAS.480.1263H}. On the other hand, in the case of slow rotators ($\vsini\sim 30~–~50\,\kms$), all fundamental key parameters, except $i$ (i.e. $R_{\rm eq}$, $\vsini$, and $\rm PA_{\rm rot}$) could be determined with acceptable uncertainties (1-2$\sigma$) by our method.
Indeed, without a rapid rotation, there is no strong gravity darkening (von Zeipel) effect, and we can not estimate separately $V_{\rm eq}$ and $i$ from the $\vsini$. Therefore, the only solution to deduce the exact flatness shape from slow rotators is to use a dense $(u,v)$ coverage with the differential visibility \citep[which contains angular size information, as done by][]{2003A&A...407L..47D}. This method could be applied to AGNs too.

We have computed the limits of accuracy of the radius $R_{\rm eq}$, inclination angle $i$ and rotation-axis position angle $\rm PA_{\rm rot}$ parameters are that we could achieve with our quality of data and this numbers are summarized in Table \ref{tab_limit} here above.
\begin{table}
\centering
\caption{Limiting and achieved accuracy for the parameter $\beta=0.165\pm0.009$ of Regulus with data used in this study \citep[][more details in Sec.~\ref{AnnexC}]{2018MNRAS.480.1263H}. The ``best possible accuracy'' is given for the estimation of a parameter from the $\vec{\epsilon}$ alone or with the spectrum $s(\lambda)$, when we assume that all other parameters are known.} \label{tab_limit}
\begin{tabular}{ccc}
\hline \hline
Parameter & Best possible & Accuracy from \\ 
 & accuracy from & MCMC$^{\star}$ fit of \\
 & $\vec{\epsilon}(\lambda)$ only & $\vec{\epsilon}(\lambda)$ \& $s(\lambda)$ \\

\hline \hline

$R_{\rm eq}$ & 1.25\% & 5.7\% \\
$V_{\rm eq}$ & $50\,\kms$ & $22\,\kms$ \\
$i$ & 2.5$^\circ$ & 6$^\circ$ \\
$\rm PA_{\rm rot}$ & 1.4$^\circ$ & 1.8$^\circ$ \\

\hline \hline

\end{tabular}
\begin{threeparttable}
\begin{tablenotes}
$(\star)$ The accuracy of MCMC fit (for Markov Chain Monte Carlo method) is that we found from our previous paper of \cite{2018MNRAS.480.1263H},
\end{tablenotes}
\end{threeparttable}
\end{table}
We conclude that SNR is insufficient to give a significant direct constraint on $\beta$ from a fit of our data. To hope to constrain $\beta$, we need at least an SNR 10 to 15 times better.

Finally, we have defined a criterion $\Gamma_\epsilon$ which helps to determine the application limits of the photocentre displacement method on fast rotating stars with gravity darkening effect. We demonstrated that for $\Gamma_\epsilon\leq 0.32$ (which corresponds to unresolved objects $V \geq 0.99$) this method is applicable, no matter the value of $\Delta \rm PA$ for couple baselines combination/averaging, and in condition of perpendicular baselines ($\Delta \rm PA\sim 90$) for single photocentre projections.  Beyond, i.e. $\Gamma_\epsilon > 0.32$, only $\phidiff$ data are usable. Also, we proved using observed spectro-interferometric data \citep[under certain conditions summarized in Sec.~\ref{AnnexC}, and applied previously by us in][]{2018MNRAS.480.1263H}, that the application field of this method can be extended to marginally resolved stars, i.e. $V\sim 0.8$, which corresponds to $\Gamma_{\epsilon}\sim 0.64$). No matter the value of $\Delta \rm PA$ for couple baselines combination/averaging and when $60^\circ\leq\Delta \rm PA\leq 90^\circ$ for single photocentre projections, especially for edge-on stars. The only relevant incompatibility that we noticed, with respect the literature results, corresponds to the inclination angle $i$ (south pole apparent instead of the north one).
Thus, to say that all key parameter information are entirely contained in the vectorial photocentre displacement, which can be measured as soon as there are two non-collinear bases is completely justified, even with a relatively bad SNR. On the other hand, the use differential phase is suitable only for $\Gamma_\epsilon\geq 0.64$ with uncertainties, (when the object is resolved; i.e. $V \leq 0.99$).

Therefore, the estimate of $\rm PA_{\rm rot}$ is possible with both methods, which opens a wide perspectives of studies, such as the determination of the ecliptic plane of targets and thus properly characterize and predict the evolution trajectory of the thousands of extra solar planets that we discover from radial-velocimetry and occultation observations, as well as the dynamical study of clusters evolution by comparing the $\rm PA_{\rm rot}$  of their star.

\appendix

\section{modeling fast rotators with SCIROCCO}
\label{AnnexAlpha}
\begin{figure}
\centering
 \includegraphics[width=1.\hsize,draft=false]{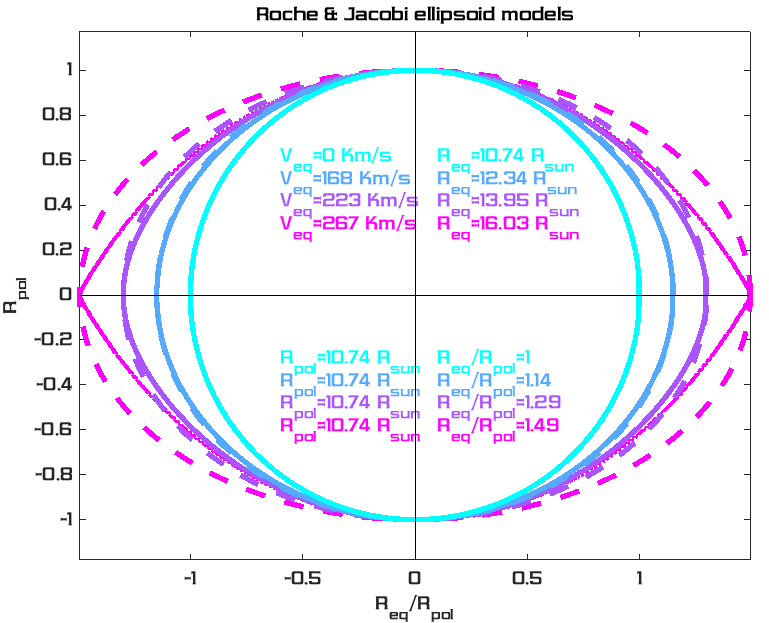} 
 \caption{Drawing of both Roche (continuous line) and ellipsoidal models (discontinuous line), for our reference Achernar-like model, with 4 different values of equatorial rotation velocities, and their corresponding equatorial radius $R_{\rm eq}$; polar radius $R_{\rm p}$ and equatorial-to-polar radii $R_{\rm eq}/R_{\rm p}$. The closer $V_{\rm eq}$ to $V_{\rm crit}=269\,\kms$ the more obvious the difference in shape between the two models, while the radii $R_{\rm eq}$ and $R_{\rm p}$ remain exactly the same.} \label{comp_roch-ellip}
\end{figure}

In contrast to the CHARRON code \citep{2012sf2a.conf..321D}, which uses the pure Roche model to compute the shape of the rotating star, the SCIROCCO code uses the simple Jacobi ellipsoidal model, as the HDUST code does \citep[][]{2006ApJ...639.1081C}, which makes the two last codes much lighter and faster. Figure \ref{comp_roch-ellip} shows the differences between the Roche and the ellipsoidal models. Although replacing the Roche model with that of the Jacobi ellipsoid does not seem to be theoretically the best approach for early-type stars, which are all centrally condensed, the angular size results of the two approaches remain the same in the limits of uncertainties (e.g. The results on the same observed VLTI/AMBER data of Achernar using CHARRON by \citealp{2012A&A...545A.130D} and through SCIROCCO by \citealp{2014A&A...569A..45H}).

SCIROCCO uses the pseudo-Voigt analytic function with 6 parameters (PVoigt$_{6}$). The mathematical function that best fits the absorption line profiles is given by :

\begin{equation}\label{pseudo-voigt6_eq}
\begin{array}{l}
\rm PVoigt_{6}(\lambda,\theta)=a_6(\theta)(L(\lambda,\theta)+(1-a_6(\theta)) G(\lambda,\theta)),
\end{array}
\end{equation}

where $\theta$ is the co-latitude, and $G$ and $L$ are the Gaussian and Lorentzian functions respectively, given by :

\begin{subeqnarray}\label{pseudo-voigt6_eq+}
& G(\lambda,\theta)=1+a_5(\theta)-a_1(\theta)\exp\Big(-\pi\frac{a_1^2(\theta)}{a_2^2(\theta)}(\lambda-\lambda_{\rm c})^2\Big),  \\
& L(\lambda,\theta)=1+a_5(\theta)-a_3(\theta)\left[1+\pi\frac{a_3^2(\theta)}{a_4^2(\theta)}(\lambda-\lambda_{\rm c})^2\right]^{-1}.
\end{subeqnarray}

The coefficients $a_1$ to $a_5$ are those of the Gaussian and the Lorentzian functions respectively (absorption lines), $a_6$ is the coefficient of the pseudo-Voigt function, and $\lambda_{\rm c}$ is the the central wavelength of the spectral line.

\begin{figure}
\centering
 \includegraphics[width=1.\hsize,draft=false]{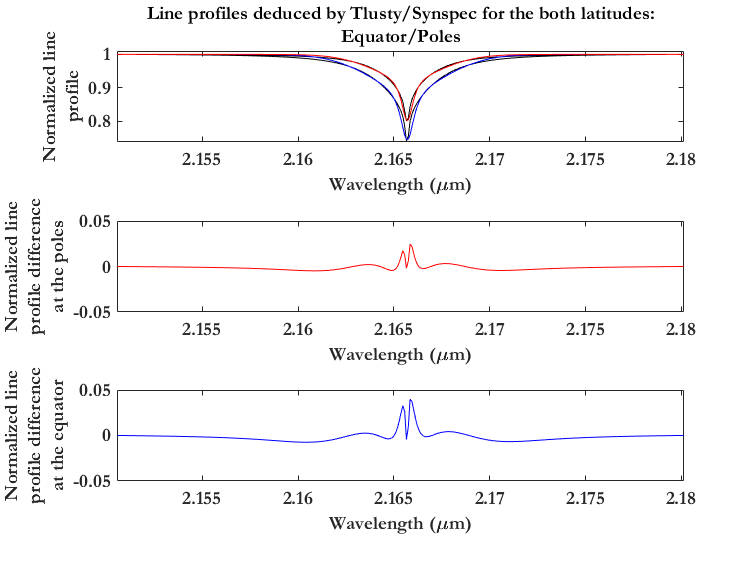}
 \caption{\textbf{Top}: PVoigt$_{6}$ functions fitted on the KURUCZ/SYNSPEC model (black line) for the equatorial line profile $[T_{\rm eff},log\ g]=[10500\,K,30\,cm.s^{-1}]$ (blue line) and the polar profile $[T_{\rm eff},log\ g]=[19000\,K,35\,cm.s^{-1}]$ (red line). \textbf{Middle:} Difference between the KURUCZ/SYNSPEC and the PVoigt$_{6}$ models at the poles. \textbf{Bottom:} The same at the equator.}\label{pseudo-voigt6}
\end{figure}

Figure \ref{pseudo-voigt6} shows the results of the numerical fit of the 6-parameter pseudo-Voigt function on the SYNSPEC profile obtained with the KURUCZ photospheric model for 2 line profiles (at the equator and at the poles). The great advantage of this method is to get very good line profiles for any wavelength with no need of interpolation, decreasing notably the computing time.
Note that the photocentre displacement is roughly proportional to the line-to-continuum amplitude ratio \citep{1995A&AS..109..401C}. An error of $1\%$ in the line profile corresponds to an error of $1\%$ in the photocenter displacement, which is much lower than the measurement uncertainty.

\section{Limits of photocentre displacements}
\label{AnnexA}
We gather here all additional figures discussed in Sec.~\ref{photocentres}.
\begin{figure*}
\centering
\includegraphics[width=0.48\hsize,draft=false]{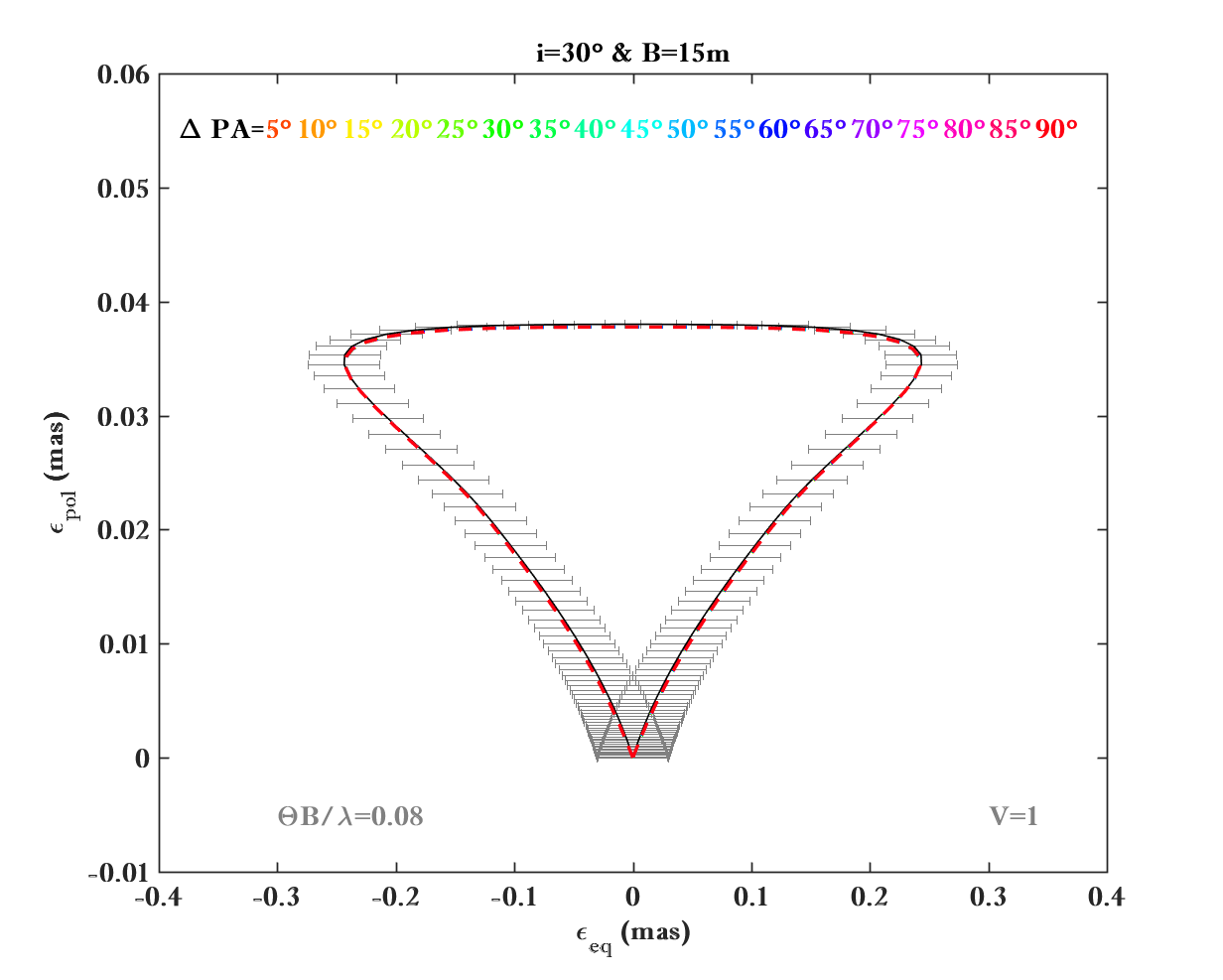}
\includegraphics[width=0.48\hsize,draft=false]{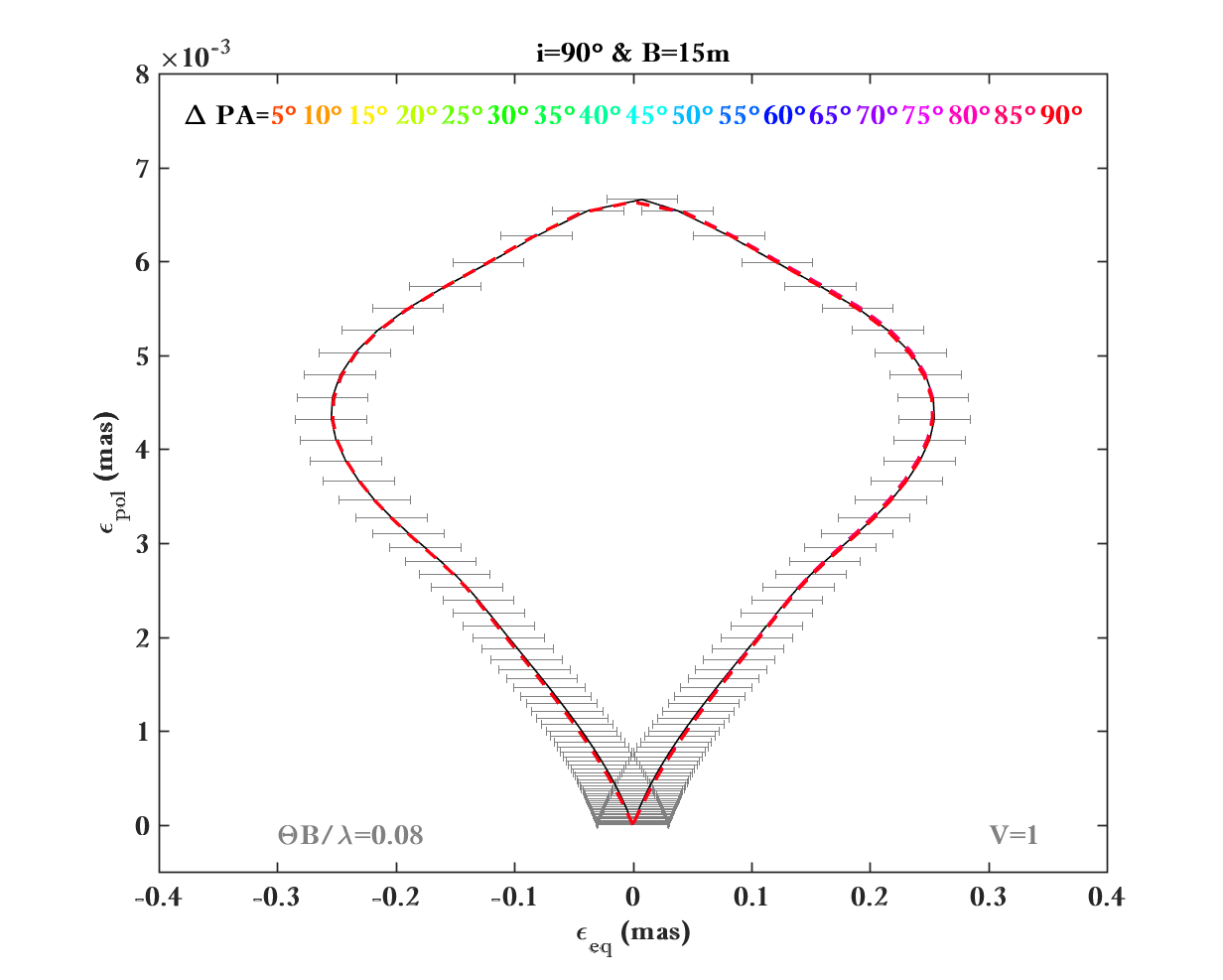}
\includegraphics[width=0.48\hsize,draft=false]{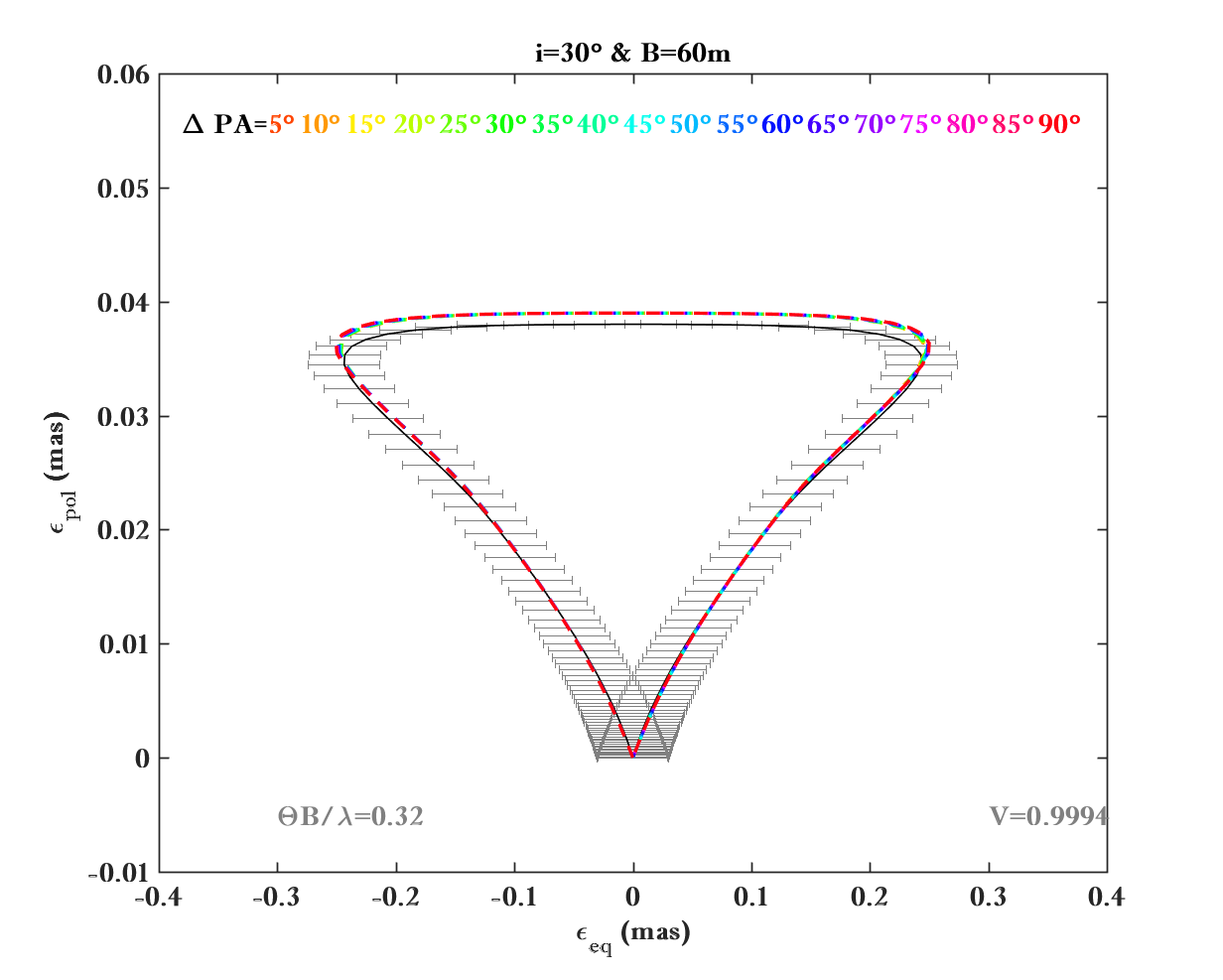}
\includegraphics[width=0.48\hsize,draft=false]{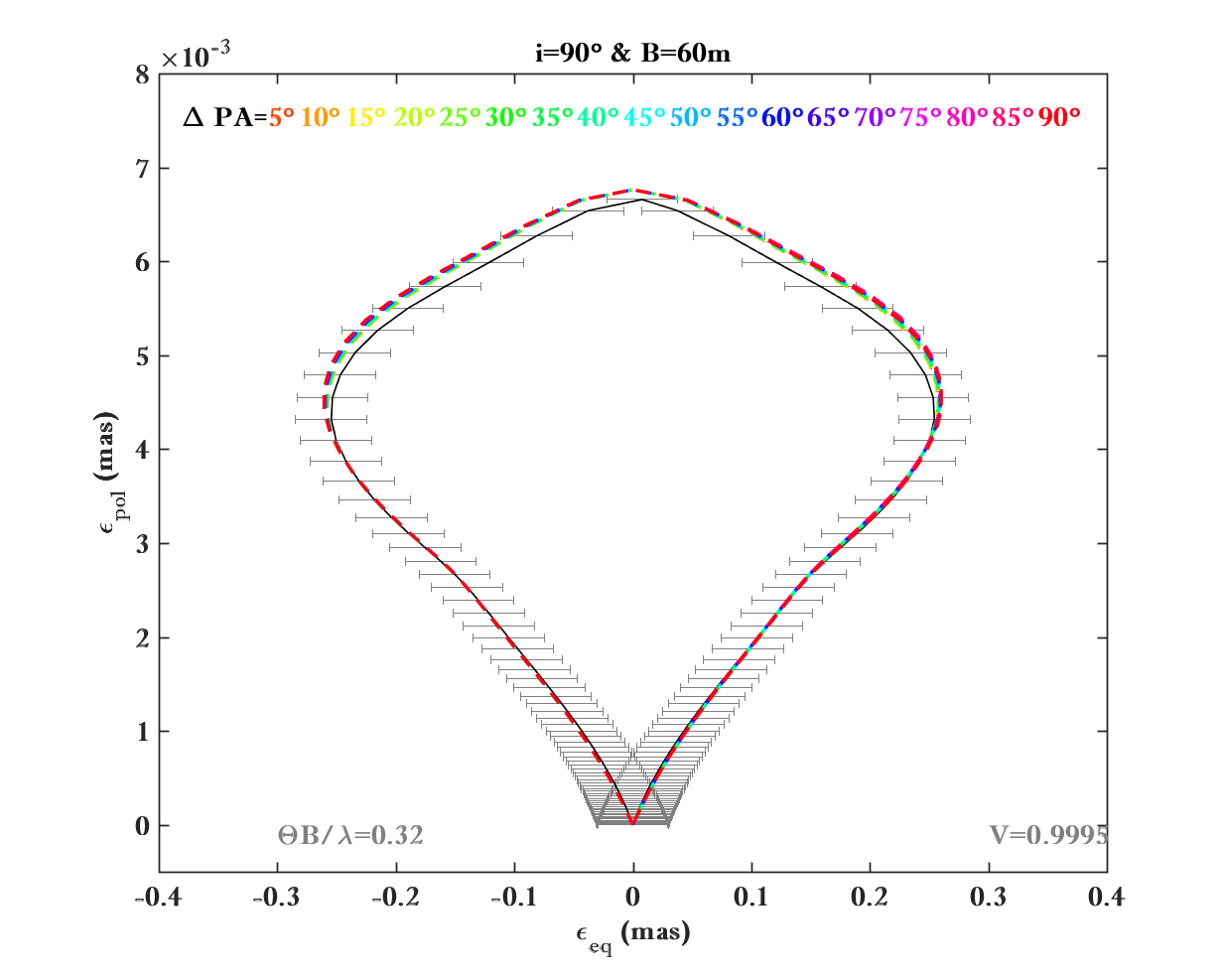}
\includegraphics[width=0.48\hsize,draft=false]{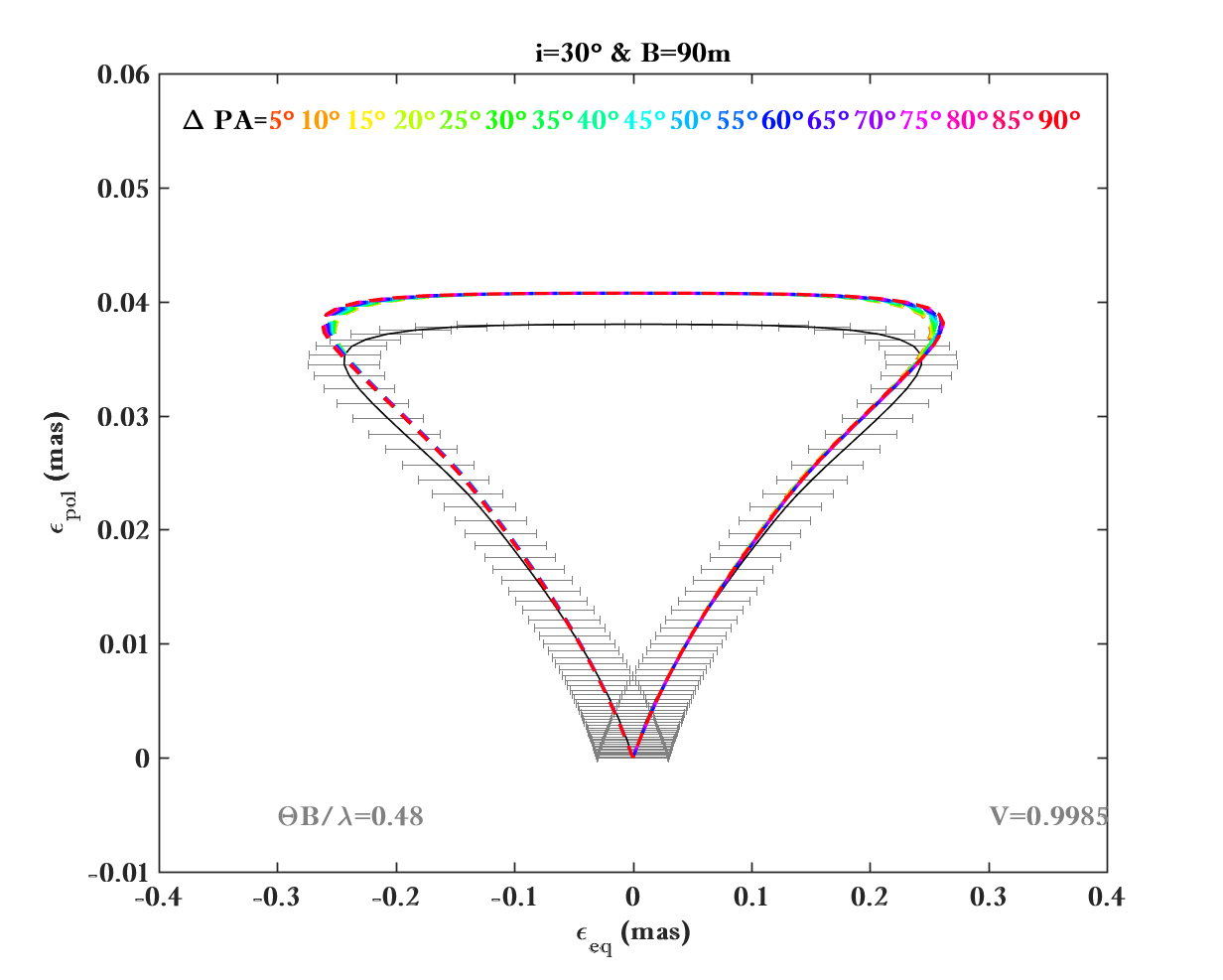}
\includegraphics[width=0.48\hsize,draft=false]{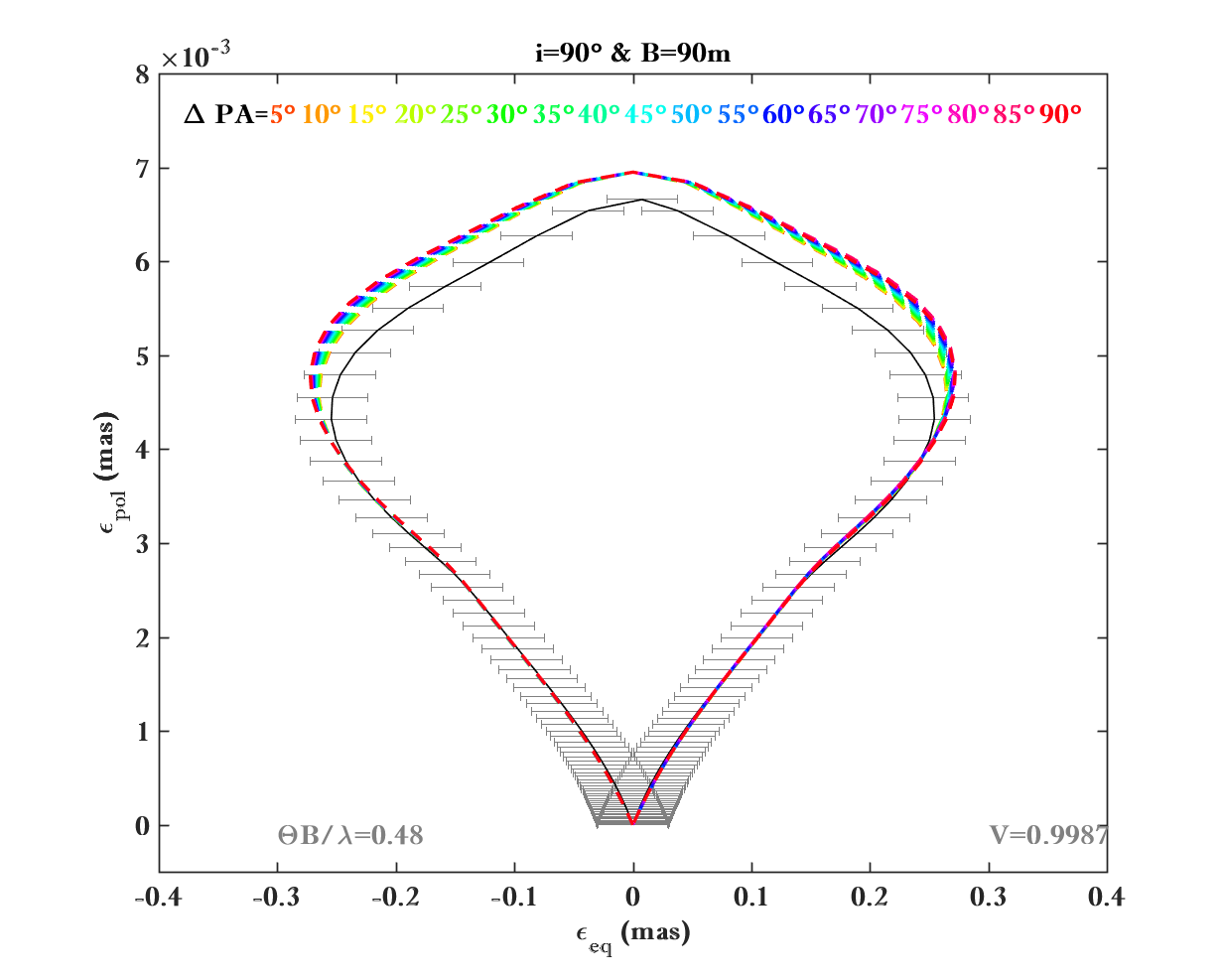}
\caption{Vectorial photocentre displacement, for a star similar to Achernar (with $\rm PA_{\rm rot}=0^\circ$ and $\beta=0.25$), around Br$\gamma$ (K-band), for various inclination angles $i$ from $30^\circ$ to $90^\circ$, and several baseline configurations $B$ from $15$ to  $90\,\rm m$, and different values of projection angle differences ($\Delta \rm PA$ from $5^\circ$ to $90^\circ$ in different colours). The absolute vectorial photocentre $E_{\rm eq,pol}$ is represented in black line. In each panel, we indicate the value of our criterion $\Gamma_\epsilon=\frac{\Theta B}{\lambda}$. The equality $E_{\rm eq,pol}\approx\epsilon_{\rm eq,pol}$ is clearly satisfied for $\Gamma_\epsilon\leq 0.32$, independently from $\Delta \rm PA$-value. The grey error bars corresponds to the measurement error of the photocentre displacements $\sim30\,\mu \rm as$, as deduced before in a previous work \citep{2018MNRAS.480.1263H}. V-value on each panel represents the average visibility over all PAs.}
\label{A8a}
\end{figure*}
\begin{figure*}
\centering
\includegraphics[width=0.48\hsize,draft=false]{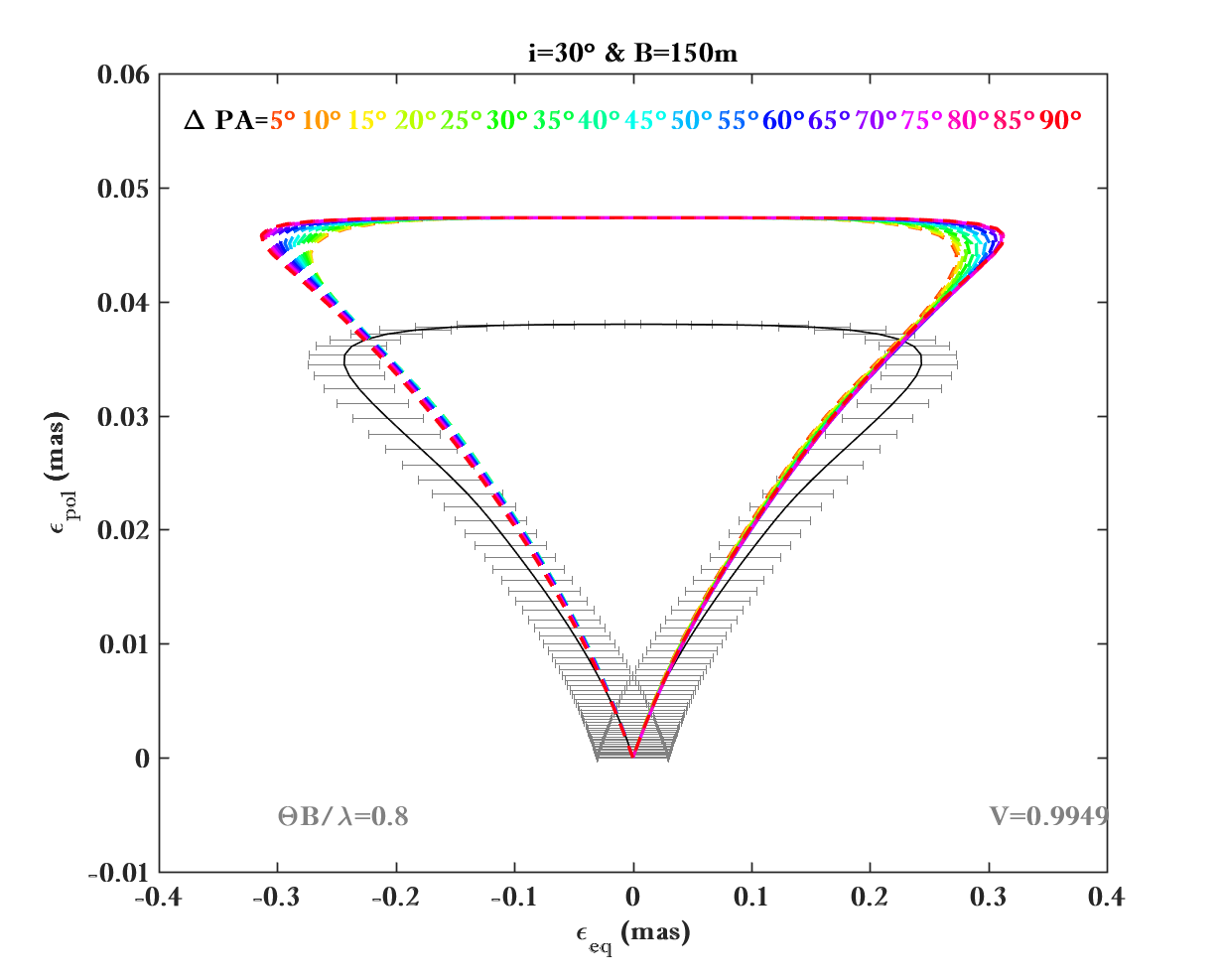}
\includegraphics[width=0.48\hsize,draft=false]{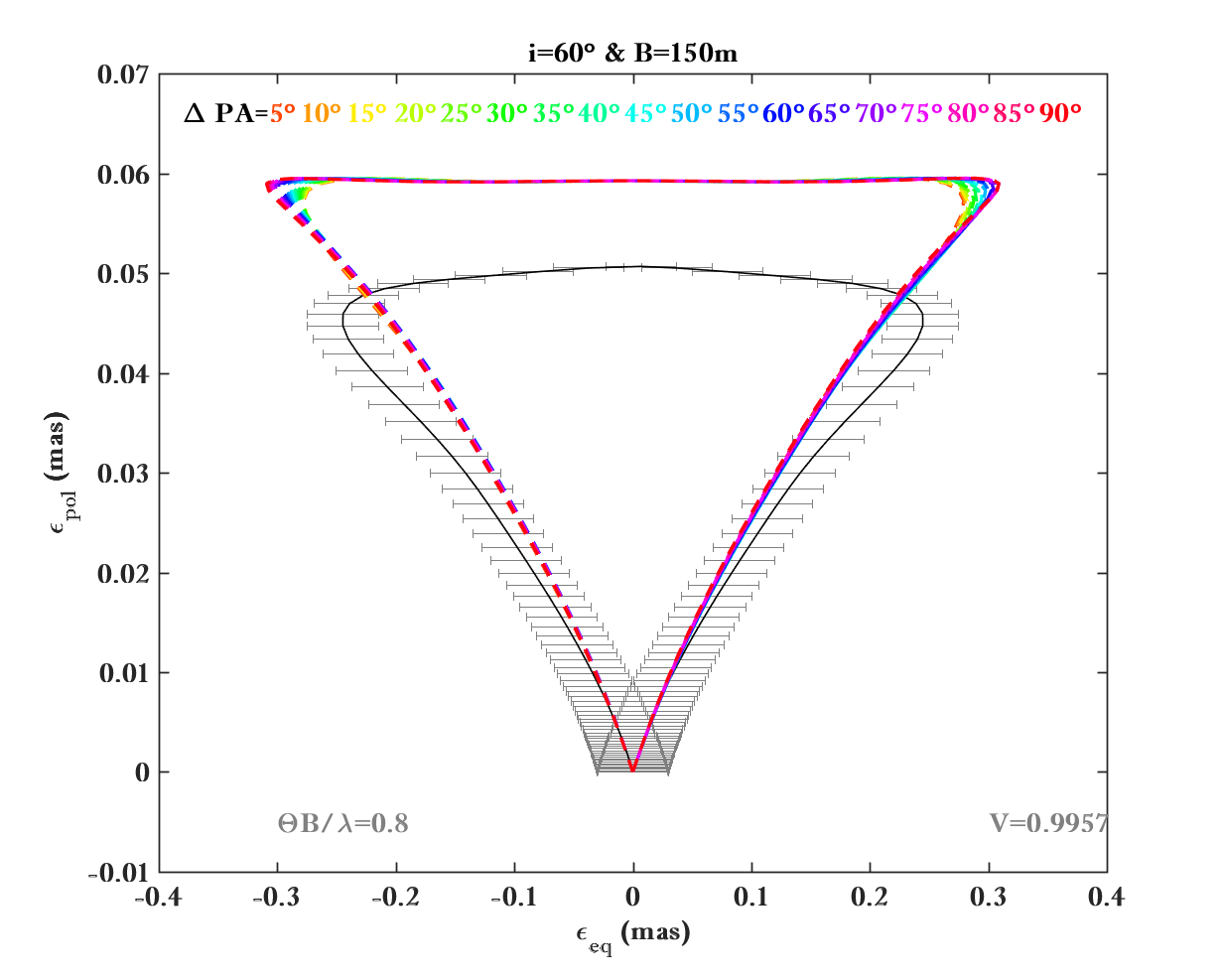}
\includegraphics[width=0.48\hsize,draft=false]{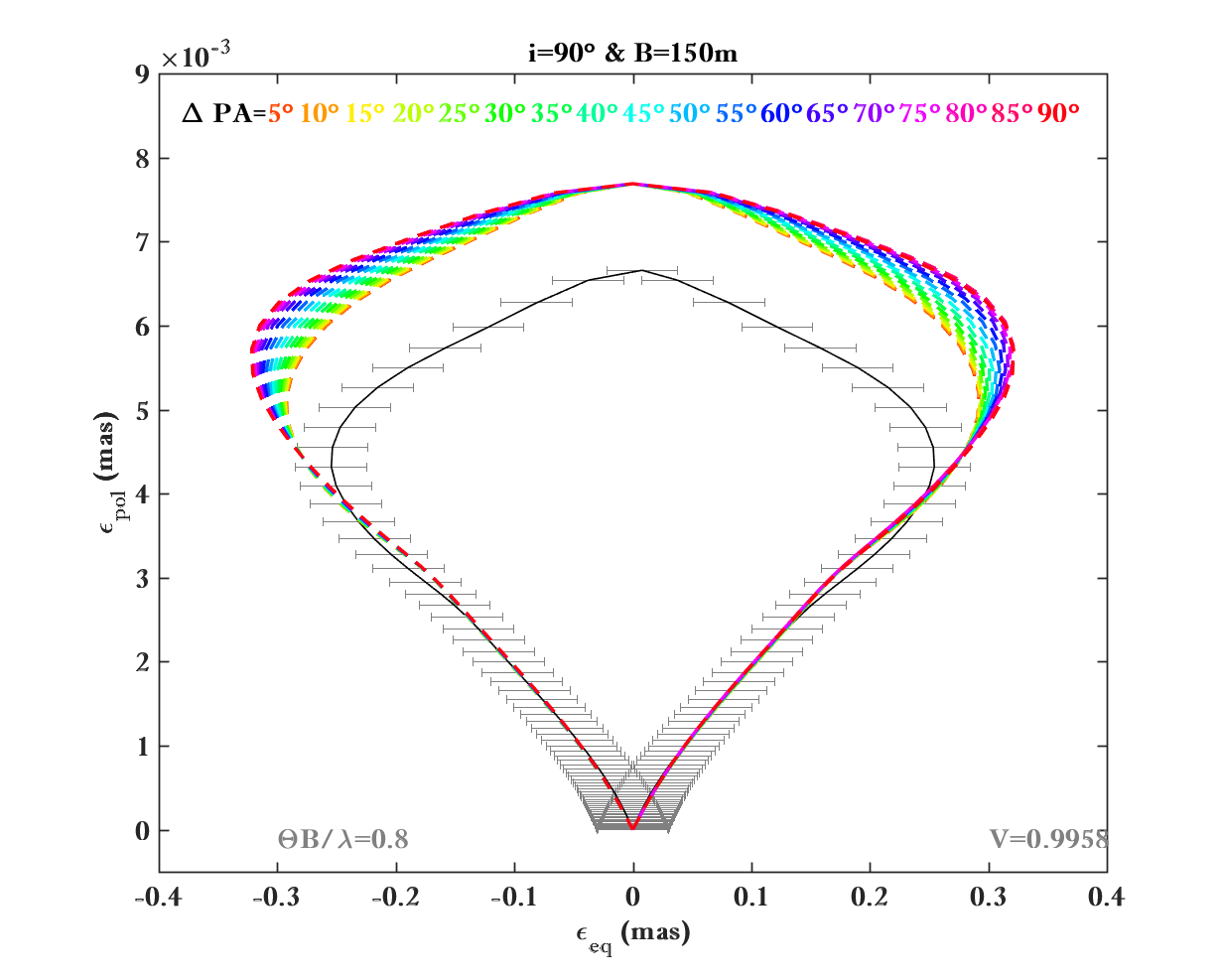}
\caption{Same as Fig.~\ref{A8a}, but for a fixed baseline length ($B=150\,\rm m$) and three different inclination angle values, namely: $i=30^\circ$, $60^\circ$, and $90^\circ$.}
\label{A8d}
\end{figure*}

\section{Dependence on physical parameters}
\label{AnnexB}
We gather here all figures discussed in Sec.~\ref{phot-sensitivity}.
We observe that the interferometric photocentre displacements (in coulour lines) fit the astrometric ones (in black line) within the average error bars of $\sigma_{\epsilon}=\pm30\,\mu as$, only for $\Gamma_\epsilon\leq 0.32$.  

\begin{figure*}
\centering
\includegraphics[width=0.48\hsize,draft=false]{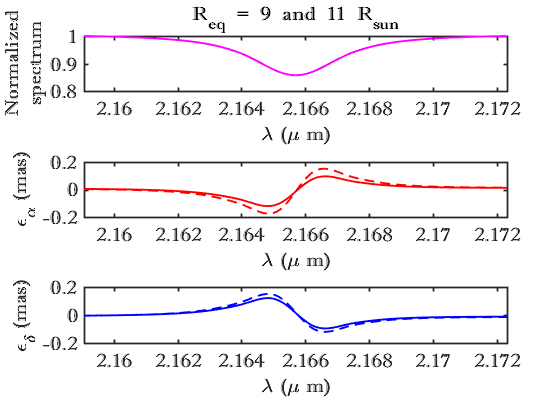}
\includegraphics[width=0.48\hsize,draft=false]{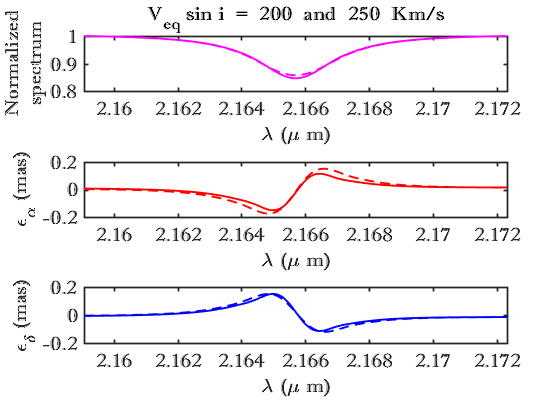}
\includegraphics[width=0.48\hsize,draft=false]{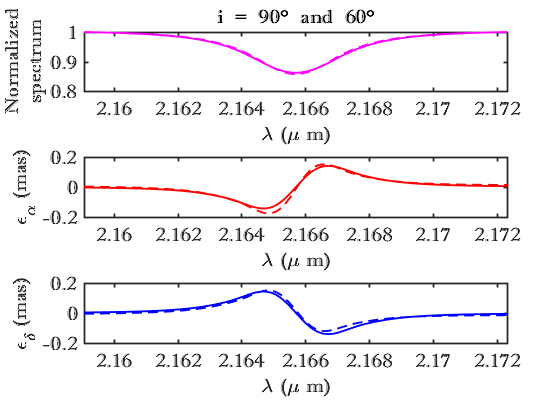}
\includegraphics[width=0.48\hsize,draft=false]{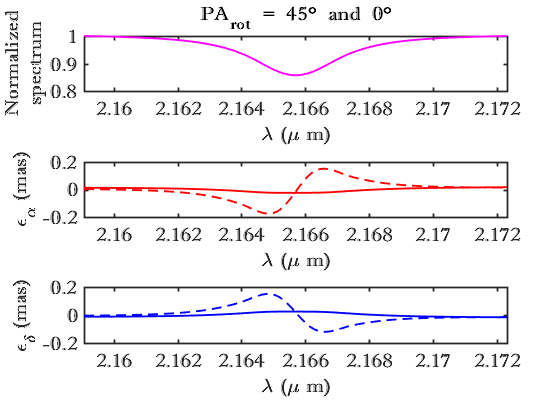}
\includegraphics[width=0.48\hsize,draft=false]{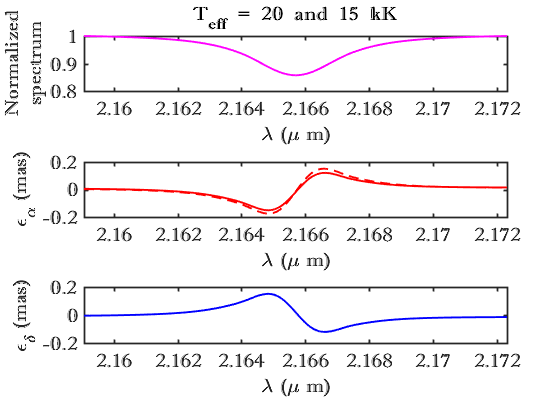}
\includegraphics[width=0.48\hsize,draft=false]{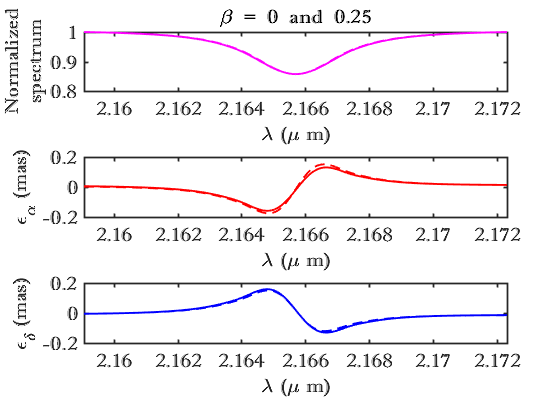}
\caption{\textbf{Top}: Dependence of simulated spectrum and perpendicular photocentres displacements ($\epsilon_\alpha$ and $\epsilon_\delta$) on $R_{\rm eq}$, $\vsini$, $i$, $PA_{\rm rot}$, $\Tmean$ and $\beta$ (solid line: tested model; dashed line: reference model as described above).} \label{fig_model1}
\end{figure*}

Fig.~\ref{fig_model1} (top) shows the dependence of simulated spectrum and perpendicular photocentres displacements ($\epsilon_\alpha$ and $\epsilon_\delta$) on $R_{\rm eq}$ and $\vsini$ (solid line: tested model; dashed line: reference model as described above). The parameters of the tested model are identical to those of the reference model, except that $R_{\rm eq} = 9\,\Rsun$. The values of the tested and reference models differ by $(\Delta_\epsilon=26\%,\Delta_s=0\%)$ for $R_{\rm eq}$. All the models were calculated across the $Br_\gamma$ line with a spectral resolution of $12000$ at four projected baselines ($B_{\rm proj} = 75\,m$ and $150\,m$, and $\rm PA = 45^\circ$ and $90^\circ$), which are typical values attained with VLTI/AMBER. These wavelengths and baselines result in visibility amplitudes between $0.6$ and $0.9$ for the studied stellar models, corresponding to a partially resolved star. While $\vsini$ changed from $250\,\kms$ to $200\,\kms$. The values of the tested and reference models differ by $(\Delta_\epsilon=23\%,\Delta_s=1.3\%)$ for $\vsini$. Fig.~\ref{fig_model1} (middle) is similar to the upper figures but for the dependence of spectrum and perpendicular photocentres displacements ($\epsilon_\alpha$ and $\epsilon_\delta$) on $i$ and $\Tmean$. $i$ changed from $60^\circ$ to $90^\circ$ and $\rm PA_{\rm rot}$ from $0^\circ$ to $45^\circ$. The values of the tested and reference models differ by $(\Delta_\epsilon=9\%,\Delta_s=0.5\%)$ for $\i$ and by $(\Delta_\epsilon=95\%,\Delta_s=0\%)$ for $\rm PA_{\rm rot}$. Fig.~\ref{fig_model1} (bottom) is similar to the upper figures but for the dependence of spectrum and perpendicular photocentres displacements ($\epsilon_\alpha$ and $\epsilon_\delta$) on $\Tmean$ and $\beta$. $\Tmean$ changed from $15000\,\rm K$ to $20000\,\rm K$ and $\beta$ from $0$ to $0.25$. The values of the tested and reference models differ by $(\Delta_\epsilon=9\%,\Delta_s=0.5\%)$ for both $\Tmean$ and $\beta$ .

\begin{figure*}
\centering
\includegraphics[width=0.48\hsize,draft=false]{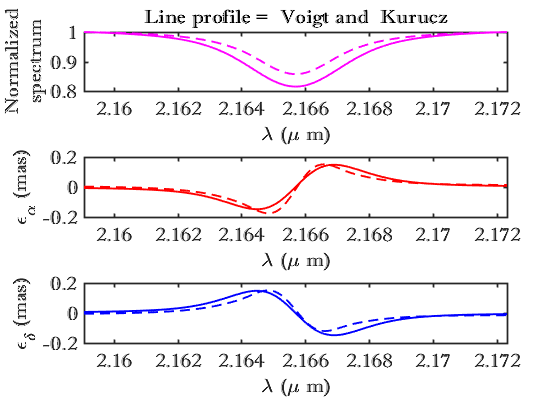}
\includegraphics[width=0.48\hsize,draft=false]{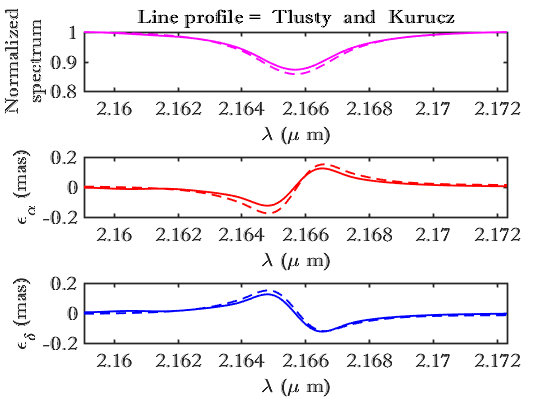}
\includegraphics[width=0.48\hsize,draft=false]{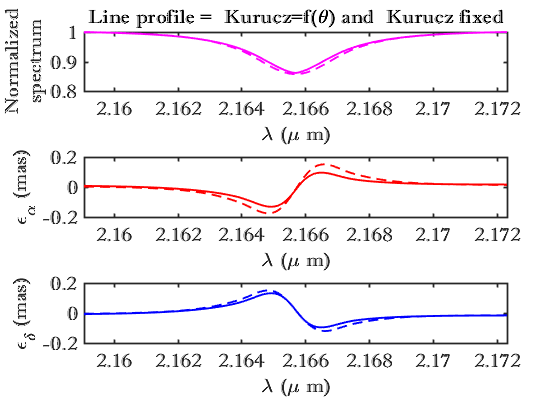}
\includegraphics[width=0.48\hsize,draft=false]{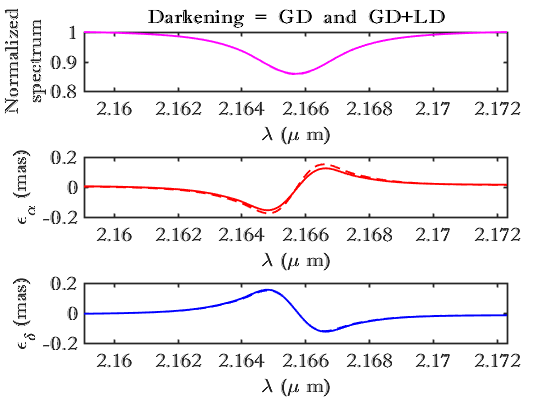}
\caption{\textbf{Top}: Similar to Fig. \ref{fig_model1} but for the dependence of simulated spectrum and perpendicular photocentres displacements ($\epsilon_\alpha$ and $\epsilon_\delta$) with different line profile sets and on the ``darkening type effect''.} \label{fig_model2}
\end{figure*}

Fig.~\ref{fig_model2} (top) is similar to Fig.~\ref{fig_model1} but for the dependence of simulated spectrum and perpendicular photocentres displacements ($\epsilon_\alpha$ and $\epsilon_\delta$) with different line profile sets. On the left panel it shown the dependence on``the line profile type'', where the solid line is related to line profile obtained with Voigt line profile; and the dashed line to the line profile obtained with Kurucz/Synspec. On right panel the solid line represents the line profile obtained with Tlusty/Synspec, and the dashed line represents the line profile obtained with Kurucz/Synspec.
The values of the tested and reference models differ by $(\Delta_\epsilon=36\%,\Delta_s=5\%)$ for first and by $(\Delta_\epsilon=15\%,\Delta_s=2\%)$ for the second one.
Fig.~\ref{fig_model2} (bottom-left) is similar to the upper figures but for the dependence on``the line profile type'' in function of the co-latitude. The solid line represents the line profile obtained with Krucz/Synspec line profile varying according to the co-latitude, and the dashed line the line profile obtained with fixed Kurucz/Synspec line profile (considering the average $[T_{\rm eff},\log g]$ of the star). While Fig.~\ref{fig_model2} (bottom-right) depicts the dependence on ``darkening type effect'' (solid line: just gravity darkening effect -no limb darkening LD effect-; dashed line: reference model as described above -gravity darkening and limb darkening GD+LD effects-).
The values of the tested and reference models differ by $(\Delta_\epsilon=23\%,\Delta_s=0.5\%)$ for first and by $(\Delta_\epsilon=8\%,\Delta_s=0.15\%)$ for the second one.
Spectrum, $\epsilon_\alpha$ \& $\epsilon_\delta$ in $Br_\gamma$  are not strongly sensitive to the darkening effect but are more sensitive to the analytic line profile used.

\section*{ACKNOWLEDGEMENTS}
\label{acknowledgements}
The first author acknowledges the support from the scientific french association PSTJ \footnote{http://www.pstj.fr/} for its official host agreement, the Lagrange and OCA for compute servers support. A special thanks to the project's grant ALMA-CONICYT N$^\circ$ 31150002 and the PI Keiichi Ohnaka who supported this work. Special thanks go to the project's grant ESO-MIXTO 2019, as well as the grants from the Fizeau European interferometry initiative (I2E). This work is also sponsored by the Chinese Academy of Sciences (CAS), through a grant to the CAS South America Center for Astronomy (CASSACA) in Santiago, Chile. WW is supported by the National Key RD Program of China No. 2019YFA0405102, the National Natural Science Foundation of China (NSFC) grant No. 42075123.


\section*{Data Availability Statements}
\label{data_availability_statements}
All observed and reduced data used in the current paper are available in electronic form at the CDS via \url{https://cdsarc.unistra.fr/ftp/J/MNRAS/511/4724/} or through VizieR via \url{http://vizier.u-strasbg.fr/viz-bin/VizieR?-source=J/MNRAS/511/4724}.

\bibliographystyle{mnras}
\bibliography{Biblio_Reg}

\begin{thebibliography}{}
\makeatletter
\relax
\def\mn@urlcharsother{\let\do\@makeother \do\$\do\&\do\#\do\^\do\_\do\%\do\~}
\def\mn@doi{\begingroup\mn@urlcharsother \@ifnextchar [ {\mn@doi@}
  {\mn@doi@[]}}
\def\mn@doi@[#1]#2{\def\@tempa{#1}\ifx\@tempa\@empty \href
  {http://dx.doi.org/#2} {doi:#2}\else \href {http://dx.doi.org/#2} {#1}\fi
  \endgroup}
\def\mn@eprint#1#2{\mn@eprint@#1:#2::\@nil}
\def\mn@eprint@arXiv#1{\href {http://arxiv.org/abs/#1} {{\tt arXiv:#1}}}
\def\mn@eprint@dblp#1{\href {http://dblp.uni-trier.de/rec/bibtex/#1.xml}
  {dblp:#1}}
\def\mn@eprint@#1:#2:#3:#4\@nil{\def\@tempa {#1}\def\@tempb {#2}\def\@tempc
  {#3}\ifx \@tempc \@empty \let \@tempc \@tempb \let \@tempb \@tempa \fi \ifx
  \@tempb \@empty \def\@tempb {arXiv}\fi \@ifundefined
  {mn@eprint@\@tempb}{\@tempb:\@tempc}{\expandafter \expandafter \csname
  mn@eprint@\@tempb\endcsname \expandafter{\@tempc}}}

\bibitem[\protect\citeauthoryear{{Aufdenberg} et~al.,}{{Aufdenberg}
  et~al.}{2006}]{2006ApJ...645..664A}
{Aufdenberg} J.~P.,  et~al., 2006, \mn@doi [\apj] {10.1086/504149}, \href
  {http://adsabs.harvard.edu/abs/2006ApJ...645..664A} {645, 664}

\bibitem[\protect\citeauthoryear{{Beckers}}{{Beckers}}{1982}]{1982AcOpt..29..361B}
{Beckers} J.~M.,  1982, \mn@doi [Optica Acta] {10.1080/713820871}, \href
  {http://adsabs.harvard.edu/abs/1982AcOpt..29..361B} {29, 361}

\bibitem[\protect\citeauthoryear{{Brandl} et~al.,}{{Brandl}
  et~al.}{2008}]{2008SPIE.7014E..1NB}
{Brandl} B.~R.,  et~al., 2008, in {McLean} I.~S.,  {Casali} M.~M.,  eds,
  Society of Photo-Optical Instrumentation Engineers (SPIE) Conference Series
  Vol. 7014, Ground-based and Airborne Instrumentation for Astronomy II. p.
  70141N (\mn@eprint {arXiv} {0807.3271}), \mn@doi{10.1117/12.789241}

\bibitem[\protect\citeauthoryear{{Carciofi} \& {Bjorkman}}{{Carciofi} \&
  {Bjorkman}}{2006}]{2006ApJ...639.1081C}
{Carciofi} A.~C.,  {Bjorkman} J.~E.,  2006, \mn@doi [\apj] {10.1086/499483},
  \href {http://adsabs.harvard.edu/abs/2006ApJ...639.1081C} {639, 1081}

\bibitem[\protect\citeauthoryear{{Che} et~al.,}{{Che}
  et~al.}{2011}]{2011ApJ...732...68C}
{Che} X.,  et~al., 2011, \mn@doi [\apj] {10.1088/0004-637X/732/2/68}, \href
  {http://adsabs.harvard.edu/abs/2011ApJ...732...68C} {732, 68}

\bibitem[\protect\citeauthoryear{{Chelli} \& {Petrov}}{{Chelli} \&
  {Petrov}}{1995}]{1995A&AS..109..401C}
{Chelli} A.,  {Petrov} R.~G.,  1995, \aaps, \href
  {http://adsabs.harvard.edu/abs/1995A\%26AS..109..401C} {109, 401}

\bibitem[\protect\citeauthoryear{{Collins} \& {Sonneborn}}{{Collins} \&
  {Sonneborn}}{1977}]{1977ApJS...34...41C}
{Collins} II G.~W.,  {Sonneborn} G.~H.,  1977, \mn@doi [\apjs]
  {10.1086/190443}, \href {http://adsabs.harvard.edu/abs/1977ApJS...34...41C}
  {34, 41}

\bibitem[\protect\citeauthoryear{{Domiciano de Souza}, {Vakili}, {Jankov},
  {Janot-Pacheco}  \& {Abe}}{{Domiciano de Souza}
  et~al.}{2002}]{2002A&A...393..345D}
{Domiciano de Souza} A.,  {Vakili} F.,  {Jankov} S.,  {Janot-Pacheco} E.,
  {Abe} L.,  2002, \mn@doi [\aap] {10.1051/0004-6361:20021015}, \href
  {http://adsabs.harvard.edu/abs/2002A\%26A...393..345D} {393, 345}

\bibitem[\protect\citeauthoryear{{Domiciano de Souza}, {Kervella}, {Jankov},
  {Abe}, {Vakili}, {di Folco}  \& {Paresce}}{{Domiciano de Souza}
  et~al.}{2003}]{2003A&A...407L..47D}
{Domiciano de Souza} A.,  {Kervella} P.,  {Jankov} S.,  {Abe} L.,  {Vakili} F.,
   {di Folco} E.,   {Paresce} F.,  2003, \mn@doi [\aap]
  {10.1051/0004-6361:20030786}, \href
  {https://ui.adsabs.harvard.edu/abs/2003A&A...407L..47D} {407, L47}

\bibitem[\protect\citeauthoryear{{Domiciano de Souza}, {Zorec}  \&
  {Vakili}}{{Domiciano de Souza} et~al.}{2012a}]{2012sf2a.conf..321D}
{Domiciano de Souza} A.,  {Zorec} J.,   {Vakili} F.,  2012a, in {Boissier} S.,
  {de Laverny} P.,  {Nardetto} N.,  {Samadi} R.,  {Valls-Gabaud} D.,
  {Wozniak} H.,  eds, SF2A-2012: Proceedings of the Annual meeting of the
  French Society of Astronomy and Astrophysics. pp 321--324

\bibitem[\protect\citeauthoryear{{Domiciano de Souza} et~al.,}{{Domiciano de
  Souza} et~al.}{2012b}]{2012A&A...545A.130D}
{Domiciano de Souza} A.,  et~al., 2012b, \mn@doi [\aap]
  {10.1051/0004-6361/201218782}, \href
  {http://adsabs.harvard.edu/abs/2012A\%26A...545A.130D} {545, A130}

\bibitem[\protect\citeauthoryear{{Espinosa Lara} \& {Rieutord}}{{Espinosa Lara}
  \& {Rieutord}}{2013}]{2013A&A...552A..35E}
{Espinosa Lara} F.,  {Rieutord} M.,  2013, \mn@doi [\aap]
  {10.1051/0004-6361/201220844}, \href
  {https://ui.adsabs.harvard.edu/abs/2013A&A...552A..35E} {552, A35}

\bibitem[\protect\citeauthoryear{{Gravity Collaboration} et~al.,}{{Gravity
  Collaboration} et~al.}{2017}]{2017A&A...602A..94G}
{Gravity Collaboration} et~al., 2017, \mn@doi [\aap]
  {10.1051/0004-6361/201730838}, \href
  {https://ui.adsabs.harvard.edu/abs/2017A&A...602A..94G} {602, A94}

\bibitem[\protect\citeauthoryear{{Gravity Collaboration} et~al.,}{{Gravity
  Collaboration} et~al.}{2018}]{2018Natur.563..657G}
{Gravity Collaboration} et~al., 2018, \mn@doi [\nat]
  {10.1038/s41586-018-0731-9}, \href
  {https://ui.adsabs.harvard.edu/abs/2018Natur.563..657G} {563, 657}

\bibitem[\protect\citeauthoryear{{Haario}, M., {Mira}  \& {Saksman}}{{Haario}
  et~al.}{2006}]{s11222-006-9438-0}
{Haario} H.,  M. L.,  {Mira} A.,   {Saksman} E.,  2006, \mn@doi [Statistics and
  Computing] {10.1007/s11222-006-9438-0}, \href
  {http://link.springer.com/article/10.1007/s11222-006-9438-0} {16, 339}

\bibitem[\protect\citeauthoryear{Hadjara}{Hadjara}{2015}]{Massi2015}
Hadjara M.,  2015, PhD thesis, Universit\'e de Nice Sophia Antipolis

\bibitem[\protect\citeauthoryear{{Hadjara}, {Vakili}, {Domiciano de Souza},
  {Millour}  \& {Bendjoya}}{{Hadjara} et~al.}{2012}]{2012sf2a.conf..533H}
{Hadjara} M.,  {Vakili} F.,  {Domiciano de Souza} A.,  {Millour} F.,
  {Bendjoya} P.,  2012, in {Boissier} S.,  {de Laverny} P.,  {Nardetto} N.,
  {Samadi} R.,  {Valls-Gabaud} D.,   {Wozniak} H.,  eds, SF2A-2012: Proceedings
  of the Annual meeting of the French Society of Astronomy and Astrophysics. pp
  533--538

\bibitem[\protect\citeauthoryear{{Hadjara}, {Vakili}, {Domiciano de Souza},
  {Millour}, {Petrov}, {Jankov}  \& {Bendjoya}}{{Hadjara}
  et~al.}{2013}]{2013EAS....59..131H}
{Hadjara} M.,  {Vakili} F.,  {Domiciano de Souza} A.,  {Millour} F.,  {Petrov}
  R.,  {Jankov} S.,   {Bendjoya} P.,  2013, in {Mary} D.,  {Theys} C.,   {Aime}
  C.,  eds,  EAS Publications Series Vol. 59, EAS Publications Series. pp
  131--140, \mn@doi{10.1051/eas/1359007}

\bibitem[\protect\citeauthoryear{{Hadjara} et~al.,}{{Hadjara}
  et~al.}{2014}]{2014A&A...569A..45H}
{Hadjara} M.,  et~al., 2014, \mn@doi [\aap] {10.1051/0004-6361/201424185},
  \href {http://adsabs.harvard.edu/abs/2014A\%26A...569A..45H} {569, A45}

\bibitem[\protect\citeauthoryear{{Hadjara}, {Petrov}, {Jankov},
  {Cruzal{\`e}bes}, {Spang}  \& {Lagarde}}{{Hadjara}
  et~al.}{2018}]{2018MNRAS.480.1263H}
{Hadjara} M.,  {Petrov} R.~G.,  {Jankov} S.,  {Cruzal{\`e}bes} P.,  {Spang} A.,
    {Lagarde} S.,  2018, \mn@doi [\mnras] {10.1093/mnras/sty1893}, \href
  {http://adsabs.harvard.edu/abs/2018MNRAS.480.1263H} {480, 1263}

\bibitem[\protect\citeauthoryear{{Jankov}}{{Jankov}}{2011}]{2011SerAJ.183....1J}
{Jankov} S.,  2011, \mn@doi [Serbian Astronomical Journal]
  {10.2298/SAJ1183001J}, \href
  {http://adsabs.harvard.edu/abs/2011SerAJ.183....1J} {183, 1}

\bibitem[\protect\citeauthoryear{{Jankov}, {Vakili}, {Domiciano de Souza}  \&
  {Janot-Pacheco}}{{Jankov} et~al.}{2001}]{2001A&A...377..721J}
{Jankov} S.,  {Vakili} F.,  {Domiciano de Souza} Jr. A.,   {Janot-Pacheco} E.,
  2001, \mn@doi [\aap] {10.1051/0004-6361:20011047}, \href
  {http://adsabs.harvard.edu/abs/2001A\%26A...377..721J} {377, 721}

\bibitem[\protect\citeauthoryear{{Kervella} \& {Domiciano de Souza}}{{Kervella}
  \& {Domiciano de Souza}}{2006}]{2006A&A...453.1059K}
{Kervella} P.,  {Domiciano de Souza} A.,  2006, \mn@doi [\aap]
  {10.1051/0004-6361:20054771}, \href
  {http://adsabs.harvard.edu/abs/2006A%26A...453.1059K} {453, 1059}

\bibitem[\protect\citeauthoryear{{Kraus}}{{Kraus}}{2018}]{2018tcl..confE..29K}
{Kraus} S.,  2018, in VLT2030. ESO, Garching, Germany. p.~29,
  \mn@doi{10.5281/zenodo.1488815}

\bibitem[\protect\citeauthoryear{{Kraus} et~al.,}{{Kraus}
  et~al.}{2012}]{2012ApJ...744...19K}
{Kraus} S.,  et~al., 2012, \mn@doi [\apj] {10.1088/0004-637X/744/1/19}, \href
  {http://adsabs.harvard.edu/abs/2012ApJ...744...19K} {744, 19}

\bibitem[\protect\citeauthoryear{{Labeyrie}}{{Labeyrie}}{1975}]{1975ApJ...196L..71L}
{Labeyrie} A.,  1975, \mn@doi [\apjl] {10.1086/181747}, \href
  {http://adsabs.harvard.edu/abs/1975ApJ...196L..71L} {196, L71}

\bibitem[\protect\citeauthoryear{Lagarde}{Lagarde}{1994}]{sl94}
Lagarde S.,  1994, PhD thesis, Universit\'e de Sophia Antipolis

\bibitem[\protect\citeauthoryear{{Le Bouquin}, {Absil}, {Benisty}, {Massi},
  {M{\'e}rand}  \& {Stefl}}{{Le Bouquin} et~al.}{2009}]{2009A&A...498L..41L}
{Le Bouquin} J.-B.,  {Absil} O.,  {Benisty} M.,  {Massi} F.,  {M{\'e}rand} A.,
   {Stefl} S.,  2009, \mn@doi [\aap] {10.1051/0004-6361/200911854}, \href
  {http://adsabs.harvard.edu/abs/2009A%26A...498L..41L} {498, L41}

\bibitem[\protect\citeauthoryear{{Maeder} \& {Peytremann}}{{Maeder} \&
  {Peytremann}}{1972}]{1972A&A....21..279M}
{Maeder} A.,  {Peytremann} E.,  1972, \aap, \href
  {http://adsabs.harvard.edu/abs/1972A\%26A....21..279M} {21, 279}

\bibitem[\protect\citeauthoryear{{Maiolino} et~al.,}{{Maiolino}
  et~al.}{2013}]{2013arXiv1310.3163M}
{Maiolino} R.,  et~al., 2013, arXiv e-prints, \href
  {https://ui.adsabs.harvard.edu/abs/2013arXiv1310.3163M} {p. arXiv:1310.3163}

\bibitem[\protect\citeauthoryear{{Meynet}}{{Meynet}}{2009}]{2009LNP...765..139M}
{Meynet} G.,  2009, in {Rozelot} J.-P.,  {Neiner} C.,  eds,  Lecture Notes in
  Physics, Berlin Springer Verlag Vol. 765, The Rotation of Sun and Stars. pp
  139--169 (\mn@eprint {arXiv} {0801.2944}),
  \mn@doi{10.1007/978-3-540-87831-5_6}

\bibitem[\protect\citeauthoryear{{Peterson} et~al.,}{{Peterson}
  et~al.}{2006}]{2006Natur.440..896P}
{Peterson} D.~M.,  et~al., 2006, \mn@doi [\nat] {10.1038/nature04661}, \href
  {http://adsabs.harvard.edu/abs/2006Natur.440..896P} {440, 896}

\bibitem[\protect\citeauthoryear{{Petrov}}{{Petrov}}{1988}]{1988ESOC...29..235P}
{Petrov} R.~G.,  1988, in {Merkle} F.,  ed.,  European Southern Observatory
  Conference and Workshop Proceedings Vol. 29, European Southern Observatory
  Conference and Workshop Proceedings. ESOC, Garching, Germany. pp 235--248

\bibitem[\protect\citeauthoryear{{Petrov}}{{Petrov}}{1989}]{1989dli..conf..249P}
{Petrov} R.~G.,  1989, in {Alloin} D.~M.,  {Mariotti} J.-M.,  eds, NATO ASIC
  Proc. 274: Diffraction-Limited Imaging with Very Large Telescopes, Cargèse,
  Corsica Island.. NATO ASIC Proc, Dordrecht, Netherlands.
p.~249

\bibitem[\protect\citeauthoryear{{Petrov}}{{Petrov}}{2019}]{2019vltt.confE..37P}
{Petrov} R.,  2019, in VLT2030. ESO, Garching, Germany. p.~37,
  \mn@doi{10.5281/zenodo.3356288}

\bibitem[\protect\citeauthoryear{{Petrov} \& {Lagarde}}{{Petrov} \&
  {Lagarde}}{1992}]{1992ASPC...32..477P}
{Petrov} R.~G.,  {Lagarde} S.,  1992, in {McAlister} H.~A.,  {Hartkopf} W.~I.,
  eds,  Astronomical Society of the Pacific Conference Series Vol. 32, IAU
  Colloq. 135: Complementary Approaches to Double and Multiple Star Research.
  p.~477

\bibitem[\protect\citeauthoryear{{Petrov} et~al.,}{{Petrov}
  et~al.}{2007}]{2007A&A...464....1P}
{Petrov} R.~G.,  et~al., 2007, \mn@doi [\aap] {10.1051/0004-6361:20066496},
  \href {http://adsabs.harvard.edu/abs/2007A%26A...464....1P} {464, 1}

\bibitem[\protect\citeauthoryear{{Porter} \& {Rivinius}}{{Porter} \&
  {Rivinius}}{2003}]{2003PASP..115.1153P}
{Porter} J.~M.,  {Rivinius} T.,  2003, \mn@doi [\pasp] {10.1086/378307}, \href
  {http://adsabs.harvard.edu/abs/2003PASP..115.1153P} {115, 1153}

\bibitem[\protect\citeauthoryear{{Rieutord}, {Espinosa Lara}  \&
  {Putigny}}{{Rieutord} et~al.}{2016}]{2016JCoPh.318..277R}
{Rieutord} M.,  {Espinosa Lara} F.,   {Putigny} B.,  2016, \mn@doi [Journal of
  Computational Physics] {10.1016/j.jcp.2016.05.011}, \href
  {https://ui.adsabs.harvard.edu/abs/2016JCoPh.318..277R} {318, 277}

\bibitem[\protect\citeauthoryear{{Sigut}, {McGill}  \& {Jones}}{{Sigut}
  et~al.}{2009}]{2009ApJ...699.1973S}
{Sigut} T.~A.~A.,  {McGill} M.~A.,   {Jones} C.~E.,  2009, \mn@doi [\apj]
  {10.1088/0004-637X/699/2/1973}, \href
  {http://adsabs.harvard.edu/abs/2009ApJ...699.1973S} {699, 1973}

\bibitem[\protect\citeauthoryear{{Stee}, {Meilland}  \& {Kanaan}}{{Stee}
  et~al.}{2008}]{2008EAS....28..135S}
{Stee} P.,  {Meilland} A.,   {Kanaan} S.,  2008, in {Wolf} S.,  {Allard} F.,
  {Stee} P.,  eds,  EAS Publications Series. EAS, Château de Pizay, France
  Vol. 28, EAS Publications Series. pp 135--144, \mn@doi{10.1051/eas:0828019}

\bibitem[\protect\citeauthoryear{{von Zeipel}}{{von
  Zeipel}}{1924a}]{1924MNRAS..84..665V}
{von Zeipel} H.,  1924a, \mnras, \href
  {http://adsabs.harvard.edu/abs/1924MNRAS..84..665V} {84, 665}

\bibitem[\protect\citeauthoryear{{von Zeipel}}{{von
  Zeipel}}{1924b}]{1924MNRAS..84..684V}
{von Zeipel} H.,  1924b, \mnras, \href
  {http://adsabs.harvard.edu/abs/1924MNRAS..84..684V} {84, 684}

\makeatother
\end{thebibliography}

\end{document}